\documentclass[%
 reprint,
nofootinbib,
 amsmath,amssymb,
 aps,
]{revtex4-1}

\usepackage{graphicx}
\usepackage{dcolumn}
\usepackage{bm}
\usepackage{hyperref}
\usepackage{color}
\usepackage{physics}
\usepackage{dsfont}
\usepackage{pgf,tikz}
\usepackage{pgfplots}
\usepackage{adjustbox}
\usepackage{soul}
\pgfplotsset{compat=1.14}
\usepackage{rotating} 

\definecolor{rainbow5red}{RGB}{237, 127, 8}
\definecolor{rainbow5orange}{RGB}{255, 161, 1}
\definecolor{rainbow5yellow}{RGB}{253, 244, 153}
\definecolor{rainbow5green}{RGB}{110, 229, 80}
\definecolor{rainbow5blue}{RGB}{139, 0, 139} 
\definecolor{lightgray}{RGB}{221,221,221}

\newcommand{\ave}[1]{\langle{#1}\rangle} 
\pgfplotsset{compat=1.14}
\begin{document}

\preprint{APS/123-QED}

\title{Twist-and-store entanglement in bimodal and spin-1 Bose-Einstein condensates}

\author{Artur Niezgoda}
\address{Faculty of Physics, University of Warsaw, ul. Pasteura 5, PL-02-093 Warsaw, Poland}
\author{Emilia Witkowska}
\address{Institute of Physics, Polish Academy of Sciences, Aleja Lotnik\'{o}w 32/46, PL-02-668 Warsaw, Poland}
\author{Safoura Sadat Mirkhalaf}
\address{Institute of Physics, Polish Academy of Sciences, Aleja Lotnik\'{o}w 32/46, PL-02-668 Warsaw, Poland}

\date{\today}

\begin{abstract}
A scheme for dynamical stabilization of entanglement quantified by the quantum Fisher information is analyzed numerically and analytically for bimodal and spin-1 Bose-Einstein condensates in the context of atomic interferometry. The scheme consists of twisting dynamics followed by a single rotation of a state which limits further evolution around stable center fixed points in the mean-field phase space. The resulting level of entanglement is of the order or larger than at the moment of rotation. It is demonstrated that the readout measurement of parity quantifies the level of entanglement during entire evolution.
\end{abstract}

\maketitle

\section{Introduction}

Entanglement is a fascinating concept of quantum physics and, as already well established, a unique resource for emerging quantum technologies. 
In metrology, for example, entangled states such as squeezed states can improve the sensitivity of interferometric measurements ~\cite{PhysRevLett.96.010401,PhysRevA.82.012337, PhysRevA.85.022322} because they allow overcoming the standard quantum limit, where sensitivity scales as $\sim 1/\sqrt{N}$ for $N$ uncorrelated particles, approaching the ultimate Heisenberg limit with scaling as $\sim 1/N$. 
Initially, this concept emerged in terms of squeezing ~\cite{mandel_wolf_1995} and very recently was applied~\cite{Barsotti_2018, PhysRevLett.123.231107, PhysRevLett.123.231108} in the optical domain. Lately, it was also successfully generated and characterized in the system composed of massive particles, namely ultra-cold atoms~\cite{Pezze2016-dx}.

In general, a production of squeezed and entangled states requires inter-atomic interaction which dynamically generates non-trivial quantum correlations between atoms. The same interaction might be undesirable after reaching the required level of entanglement because it can still dynamically degrade entanglement or inter-atomic correlations. The twisting types of interaction \cite{PhysRevA.47.5138, PhysRevA.88.033629} allows a uniform description of dynamical entanglement generation for many setups composed of cold atoms~\cite{Pezze2016-dx}, e.g. for cavity induced spin squeezing~\cite{PhysRevLett.104.073604, Bohnet_2014, Haas180} and from spin-changing collisions in bimodal~\cite{Riedel2010,Gross2010,Laudat_2018} and spin-1 Bose-Einstein condensates~\cite{Hamley2012, Lucke2011, PhysRevLett.117.143004, Zou6381, qu2020probing}. In particular, in the latter setup the undesired effect of interaction is difficult to reduce. 

\begin{figure}[hbt!]
    \centering
    \includegraphics[width=1.\linewidth]{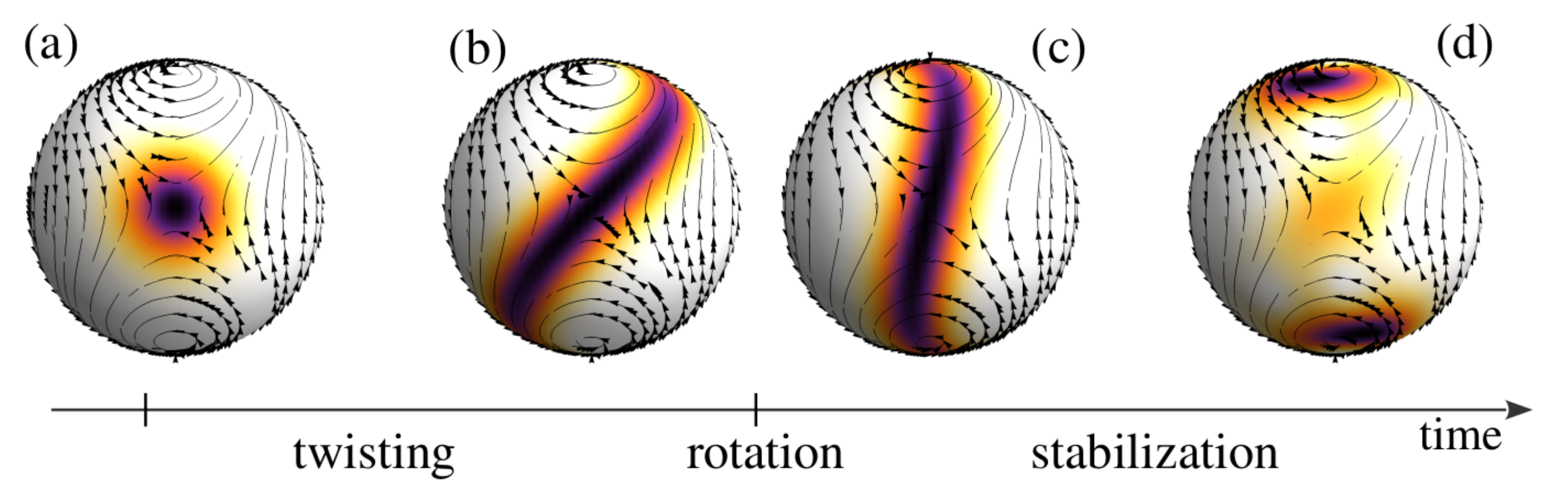}
    \caption{Illustration of the method for entanglement stabilization and storage. 
    The condensate is initialized at the unstable fixed point (a). 
    Initial evolution produces spin squeezing and entanglement along the diverging manifold of the separatrix (b). The quantum state is quickly rotated to locate it around the two stable fixed points (c). Subsequent evolution of the rotated state (d) is confined around stable fixed points leading to the stable value of the quantum Fisher information with Heisenberg scaling.}
    \label{fig:protocol}
\end{figure}

In this paper, we propose a simple method for entanglement stabilization and storage by a single rotation of a state in bimodal and spin-1 Bose-Einstein condensates. The idea is very simple and, as is illustrated in Fig.~\ref{fig:protocol}, it employs a structure of the mean-field phase space of the system Hamiltonian. The structure is the same for both bimodal and spin-1 condensates as we demonstrate in Section~\ref{sec:model}. The method considers a generalized Ramsey protocol with an additional rotation of a state applied after twisting dynamics. Once the initial spin coherent state, placed around a saddle point, is twisted along constant energy lines, the single rotation puts the state around two stable center points where further dynamics is confined and stabilized. 
We provide details of the scheme in Section~\ref{sec:protocol}. 
We observe that the value of the quantum Fisher information (QFI), which quantifies not only the level of the sensitivity of interferometric measurements but also the level of entanglement~\cite{PhysRevA.85.022321}, remains at least as at the moment of rotation, moreover it can initially grow. 
We provide an analytical explaination of this feature of the QFI using a single argument of an energy conservation in Section~\ref{sec:storage}. Therefore, we conclude that the QFI can exhibit Heisenberg scaling with the pre-factor of the order of one during the entire evolution in the idealized scheme considered in this paper. 

The best sensitivity, and therefore the QFI value, can be estimated using the signal-to-noise ratio~\cite{braunstein1994} when appropriate readout measurement is provided. In general, identification of a good observable to measure that gives the highest precision is a difficult task, in particular for non-Gaussian states. It might require measurements of higher order correlation functions~\cite{PhysRevLett.122.090503}.  Here, in Section~\ref{sec:parity}, we define the parity operators for both the bimodal and spin-1 systems. We show analytically, and confirm numerically, that the measurement of parity~\cite{parity_m_2011} allows the sensitivity to saturate the QFI value. We prove this, by using only the fact of parity conservation. The measurement can be robust against phase noise if the operator representing the noise commutes with the parity operator~\cite{PhysRevA.93.022331}.

\section{The model and structure of classical mean-field phase space}\label{sec:model}

The desired structure of the mean-field phase space is composed of two stable center fixed points located symmetrically on both sides of an unstable saddle fixed point. We concentrate here on the two systems widely explored theoretically and experimentally in the ultra-cold atomic gases, namely bimodal and spinor Bose-Einstein condensates.

\subsection{Bimodal condensate}

We consider here the twisting model enriched by a linear coupling term between the two modes $a$ and $b$ turning the state along an orthogonal direction of the form
\begin{eqnarray}\label{eq:bimodalH}
\hat{H}_{\rm BI}=\hbar \chi \hat{S}_z^2 - \hbar \Omega \hat{S}_x ,
\end{eqnarray}
where $\hat{S}_x=\frac{1}{2} \left( \hat{a}^\dagger \hat{b} + \hat{b}^\dagger \hat{a} \right)$, $\hat{S}_y=\frac{1}{2i} \left( \hat{a}^\dagger \hat{b} - \hat{b}^\dagger \hat{a} \right)$, $\hat{S}_z=\frac{1}{2} \left( \hat{a}^\dagger \hat{a} - \hat{b}^\dagger \hat{b} \right)$ are pseudo-spin operators satisfying the cyclic commutation relation $[\hat{S}_l,\hat{S}_n] =i\sum_{m} \epsilon_{lnm} \hat{S}_m$, where $\epsilon_{lnm}$ is the Levi-Civita symbol and $\hat{a} (\hat{a}^\dagger)$ and $\hat{b} (\hat{b}^\dagger)$ are bosonic mode annihilation (creation) operators of an atom in the mode $a$ ($b$). The above Hamiltonian describes two weakly-coupled Bose-Einstein condensates interacting with the strength $\chi$ in the presence of an external field of the strength~$\Omega$. 
The model can be realized experimentally employing either a double-well trapping potential~\cite{Gross2010, Trenkwalder2016} or internal (e.g. two hyperfine atomic states) degrees of freedom~\cite{Riedel2010}.

\begin{figure}[htb!]
    \centering
    \includegraphics[width=1.\linewidth]{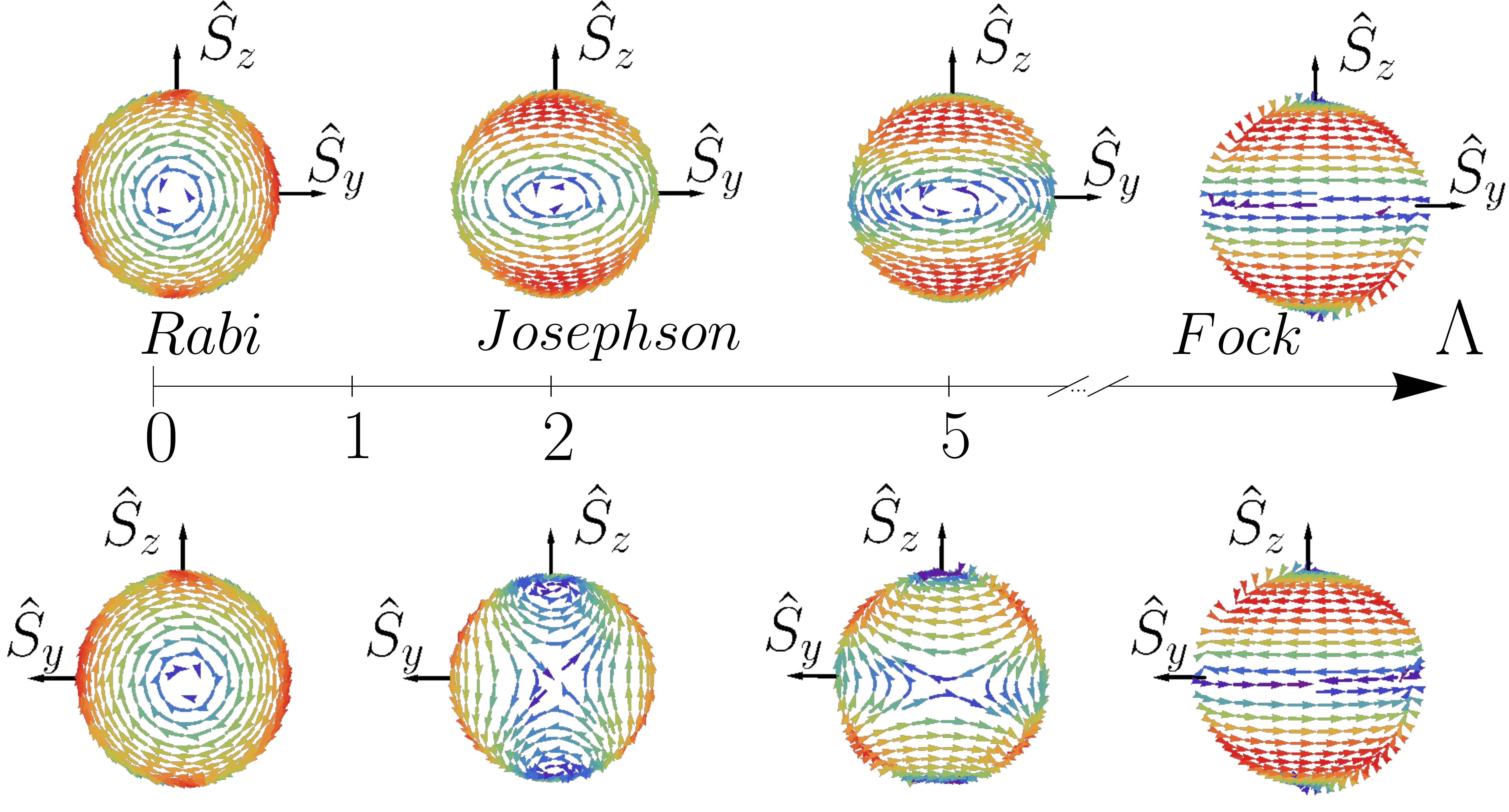}
    \caption{
    The structure of classical mean-field phase space for the bimodal system versus $\Lambda$. 
    The upper panels show the view from the positive side of $x$-axis, while the bottom panels show the view from the negative side. 
    The principal three regimes are distinguished as indicated by a name above the $\Lambda$ axis, and discussed in the main text. In this paper, we consider $\Lambda=2$ and the initial state located around an unstable fixed point located along the $x$-axis, at the negative side of it.}
    \label{fig:phase_portrait_bimodal}
\end{figure}

To obtain the mean-field phase space one can calculate an average value of (\ref{eq:bimodalH}) over the spin coherent state 
\begin{equation}\label{eq:scs_bimodal}
    |\varphi, \theta \rangle_{\rm BI}=e^{-i\varphi \hat{S}_x} e^{-i\theta \hat{S}_y}|N,0\rangle ,
\end{equation}
where $\frac{\hat{a}^\dagger{}^N}{\sqrt{N!}}|0,0\rangle=|N,0\rangle$ and $\varphi\in [0,2\pi],\, \theta\in [0,\pi]$.
The spin coherent state is a double rotation of a maximally polarized state when all atoms are in the state $a$.\footnote{Alternatively, one can substitute the quantum mechanical operators by complex numbers $\hat{a} \to \sqrt{N_a}e^{i\varphi_a}$ ($\hat{b} \to \sqrt{N_b}e^{i\varphi_b}$), where $Nz=N_a - N_b$ and $\varphi=\varphi_a - \varphi_b$ corresponds
to the relative phase between the two internal states. This procedure is not obvious for the spinor system as we will concentrate on symmetric subspace of the Hamiltonian.} This leads to
\begin{equation}\label{eq:HBI}
H_{\rm BI}=\frac{\Lambda}{2}z^2 - \sqrt{1-z^2} {\rm cos}\varphi,    
\end{equation}
where $z={\rm cos}\theta$ and $\Lambda = \chi N/\Omega$~\cite{PhysRevA.55.4318} while keeping the leading terms. 
The parameters $(z, \varphi)$ are conjugate coordinates which draw trajectories in the mean-field phase space resulting from the Hamilton equations $\dot{z}=-\sqrt{1-z^2}{\rm sin}\varphi$ and $\dot{\varphi}=\Lambda z + \frac{z}{\sqrt{1-z^2}}{\rm cos}\varphi$.
The desired by our protocol feature of the above mean-field phase space trajectories is a presence of suitable configuration of stable and unstable fixed points. The position of fixed points is a solution of $(\dot{z}=0$, $\dot{\varphi}=0)$. The resulting structure of phase space is shown in Fig.~\ref{fig:phase_portrait_bimodal}. 
The three principal regimes can be distinguished depending on the value of $\Lambda$ and characterized by different positions and number of fixed points~\cite{RevModPhys.73.307,PhysRevLett.79.4950}. 
The first one is the ``Rabi'' regime for $\Lambda < 1$ in which the linear term governs the time evolution of the system. In the limit $\Lambda \to 0$, the evolution is similar to resonant Rabi oscillations with $N$ independent particles. The two stable center fixed points are localized at $(z,\varphi)=(0,0)$ and $(z,\varphi)=(0,\pi)$. The second is the ``Josephson'' regime appearing for $\Lambda > 1$. In this regime the fixed point localized at $(z,\varphi)=(0,\pi)$ becomes unstable and the two new stable fixed points form at $(z,\varphi)=(\pm \sqrt{1-\frac{1}{\Lambda^2}},\pi)$. The change happens just after the bifurcation point at $\Lambda=1$. In this regime, the characteristic ``$\infty$'' shape is drawn up by trajectories centered around an unstable fixed point at $(z,\varphi)=(0,\pi)$, see Fig.~\ref{fig:phase_portrait_bimodal}. 
The ``$\infty$'' shape is the one that allows storing entanglement.
Finally, the third ``Fock'' regime occurs for $\Lambda \gg 1$, when the phase portrait has the same structure as  the one-axis twisting~(OAT)~model~\cite{PhysRevA.47.5138}. It is composed of two stable fixed points at $(z,\varphi)=(\pm 1, \varphi)$, and the unstable one at $(z,\varphi)=(0, \varphi)$.

\subsection{Spinor condensate}

The same structure of the mean-field phase space can be realized in spinor Bose-Einstein condensates with three internal levels instead of two, as discussed above. It can be seen in the single mode approximation (SMA) where all atoms from different Zeeman states occupy the same spatial mode $\phi(\mathbf{r})$ which satisfies the Gross-Pitaevskii equation.
The many-body Hamiltonian is expressed in terms of annihilation (creation) operators $\hat{a}_{m_F} (\hat{a}_{m_F}^\dagger)$ of an atom in the $m_F$ Zeeman state and spin-1 operators, which we collected in the vector $\vec{\Lambda}=\{\hat{J}_x, \hat{Q}_{yz}, \hat{J}_y, \hat{Q}_{zx}, \hat{D}_{xy}, \hat{Q}_{xy},\hat{Y},\hat{J}_z\}$ (see Appendix \ref{app:lambda} for definitions) is
\begin{align}\label{eq:spinorH}
\frac{\hat{H}_{\rm S}}{c_2'} = -\frac{1}{2N} \hat{J}^2 + q \hat{N}_s,
\end{align}
after dropping constant terms~\cite{KAWAGUCHI2012253, RevModPhys.85.1191}. Here, the energy unit $c'_2 = N \frac{|c_2|}{2} \int d^3r |\phi(\mathbf{r})|^4$ is associated to the spin interaction energy, $\hat{J}^2 = \hat{J}_x^2+\hat{J}_y^2 + \hat{J}_z^2$ and $\hat{N}_s=\hat{a}^\dagger_1 \hat{a}_1 + \hat{a}^\dagger_{-1} \hat{a}_{-1}$~\cite{Ho1998,Machida,PhysRevA.88.033629}. The last term in (\ref{eq:spinorH}) is due to quadratic Zeeman effect which can have contribution from the external magnetic field or microwave light field~\cite{PhysRevA.73.041602}. The value of $q$ can be either positive or negative. The Hamiltonian~\eqref{eq:spinorH} conserves the $z$-component of the collective angular momentum operator $[\hat{H}_{\rm S},\hat{J}_z] = 0$; hence, the linear Zeeman energy term is irrelevant and is omitted here. The magnetization $M \in [ -N, N]$, being the eigenvalue of the $\hat{J}_z$ operator, is a conserved quantity. The above Hamiltonian can be engineered e.g. in $F=1$ hyperfine manifold using~Rb$^{87}$~atoms~\cite{Hamley2012,Gerving2012,PhysRevLett.117.143004,Lange416,PhysRevLett.112.155304}.

\begin{figure*}
    \centering
    \includegraphics[width=1.\linewidth]{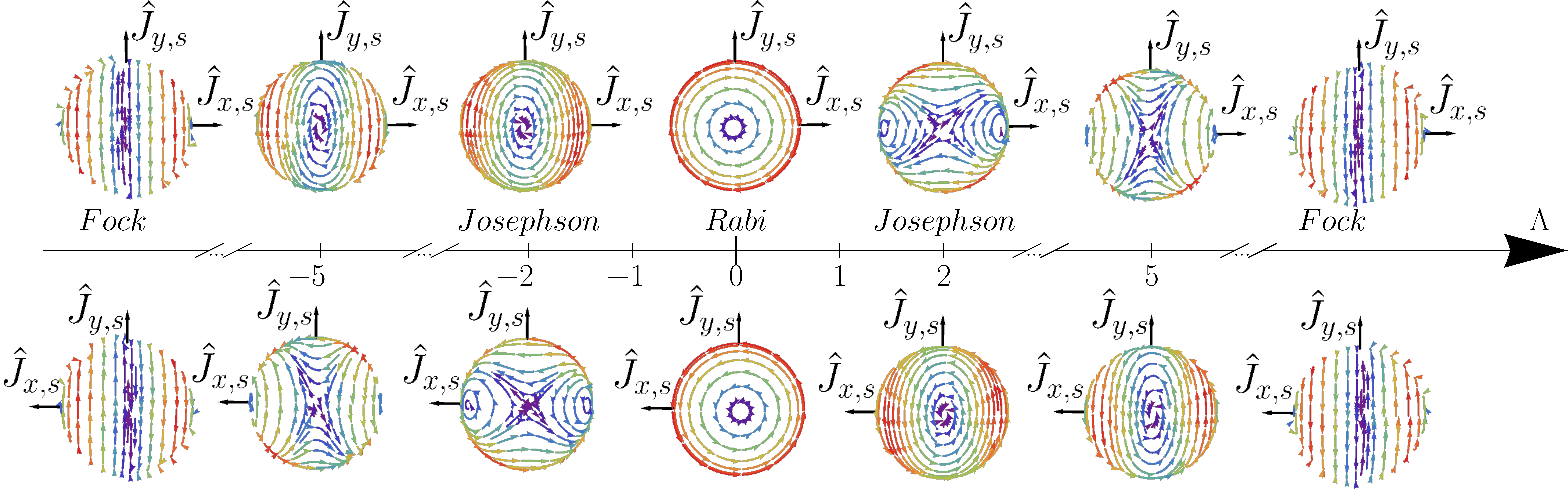}
    \caption{The structure of phase portraits of the spinor system versus $\Lambda$ in the symmetric subspace. The upper panels show a view of the north poles of the Bloch sphere, while the bottom panels show a view of south poles. The structure is the same as the one for the bimodal system, provided that the latter is rotated by $\pi/2$ around $y$-axis. The three different regimes appear as well and are indicated above the $\Lambda$ axis. In this paper, we focus on $\Lambda=-2$ and the initial state located around the unstable saddle fixed point on the south pole of the Bloch sphere.}
    \label{fig:phase_portrait_spinor}
\end{figure*}

For our purposes it is convenient to introduce the symmetric and anti-symmetric bosonic annihilation operators,
$\hat{g}_s=(\hat{a}_1+\hat{a}_{-1})/\sqrt{2}$ and $\hat{g}_a=(\hat{a}_1-\hat{a}_{-1})/\sqrt{2}$,
and the corresponding pseudo-spin operators 
\begin{align}
&\hat{J}_{x,\sigma}=\hat{a}^\dagger_0\hat{g}_\sigma+\hat{a}_0\hat{g}^\dagger_\sigma, \\
&\hat{J}_{y,\sigma}=i(\hat{a}^\dagger_0\hat{g}_\sigma-\hat{a}_0\hat{g}^\dagger_\sigma), \\
&\hat{J}_{z,\sigma}=\hat{g}_\sigma^\dagger \hat{g}_\sigma - \hat{a}^\dagger_0\hat{a}_0,
\end{align}
where indices $\sigma=s$ and $\sigma=a$ refer to symmetric and anti-symmetric subspace. The above operators have cyclic commutation relations, e.g. $[\hat{J}_{x,\sigma}, \hat{J}_{y,\sigma}]=2 i \hat{J}_{z,\sigma}$. Note, the symmetric subspace is spanned by 
$\{ \hat{J}_{x,s}, \hat{J}_{y,s}, \hat{J}_{z,s}\} = \{ \hat{J}_x, \hat{Q}_{yz}, \frac{1}{2}( \sqrt{3}\hat{Y}+\hat{D}_{xy} )\}$ while the anti-symmetric subspace by
$\{ \hat{J}_{x,a}, \hat{J}_{y,a}, \hat{J}_{z,a}\} = \{  \hat{Q}_{zx},\hat{J}_y, \frac{1}{2}( \sqrt{3}\hat{Y} - \hat{D}_{xy} )\}$. 
The spin-1 Hamiltonian~(\ref{eq:spinorH}) can be expressed in terms of symmetric and anti-symmetric operators~\cite{PhysRevA.65.033619, PhysRevA.97.032339} as 
\begin{eqnarray}
\frac{\hat{H}_{\rm S}}{|c_2'|} &=& -\frac{1}{2N} \hat{J}^2_{x,s} +\frac{q}{3}\hat{J}_{z,s}
-\frac{1}{2N} \hat{J}^2_{y,a} +\frac{q}{3}\hat{J}_{z,a} \nonumber \\
&-&\frac{1}{2N}\left(  \hat{g}^\dagger_s \hat{g}_a + \hat{g}^\dagger_{a} \hat{g}_s \right)^2
\label{eq:spinorHsymantysym}
\end{eqnarray}
up to constant terms.
The Hamiltonian \eqref{eq:spinorHsymantysym} is a sum of two (non-commuting) bimodal Hamiltonians for symmetric and anti-symmetric operators, as in (\ref{eq:bimodalH}), provided that they are rotated in respect to each other, plus a mixing term which comes from the $\hat{J}_z^2$ operator. 
Therefore, the mean-field phase space of the spinor system in each subspace is expected to have the same structure as the bimodal condensate~(\ref{eq:bimodalH}). 

To show this, we concentrate here on the symmetric subspace spanned by the symmetric pseudo-spin operators $\hat{J}_{x,s}, \hat{J}_{y,s}, \hat{J}_{z,s}$ (the anti-symmetric mean-field subspace is provided in Appendix \ref{app:meanfieldanti}). The structure of mean-field phase space can be obtained by calculating an average value of (\ref{eq:spinorH}) over the spin coherent state defined for the symmetric subspace as
\begin{equation}\label{eq:SCS_spinor}
    |\varphi, \theta \rangle_{\rm S}=e^{-i\varphi \hat{J}_{z,s}/2} e^{-i\theta \hat{J}_{y,s}/2}|N0 \rangle_s
\end{equation}
where $|N0\rangle_s = \frac{\hat{g}_s^\dagger{}^N}{\sqrt{N!}} |000\rangle $ and once again $\varphi\in (0,2\pi),\, \theta\in (0,\pi)$. The spin coherent state (\ref{eq:SCS_spinor}) can be interpreted as a double rotation of maximally polarized state $|N0\rangle_s$ in the symmetric subspace, when all atoms are in the symmetric mode.
The state $|N0\rangle_s$ is an eigenstate of $\hat{J}_{z,s}$ such that $\hat{J}_{z,s} |N0\rangle_s= N |N0\rangle_s$, and is located on the north pole of the Bloch sphere in the symmetric subspace. 
In terms of spin-1 operators it reads $|N0\rangle_s = e^{-i \pi/4 \hat{Q}_{xy}}|N00\rangle$.
On the contrary, the state with N atoms in the $m_F=0$ mode, $\frac{\hat{a}_0^\dagger{}^N}{\sqrt{N!}} |000\rangle =|0N0\rangle$, lies on the south pole of the same Bloch sphere. 
In addition, one can show that
\begin{equation}
      |\varphi, \theta \rangle_S = \frac{1}{\sqrt{N!}}\left[ \hat{g}_s^\dagger {\rm cos}\frac{\theta}{2} + \hat{a}_0^\dagger {\rm sin }\frac{\theta}{2} e^{i\varphi}\right]^N|000\rangle ,
 \end{equation}
up to the constant phase factor.
We use the above expression while illustrating an arbitrary state $|\Psi \rangle $ on the Bloch sphere in the symmetric subspace with the help of the Husimi function $Q_S (\varphi, \theta)=|\langle \Psi | \varphi, \theta \rangle_S|^2$.
 
An average value of the spin-1 Hamiltonian (\ref{eq:spinorH}) over the spin coherent state (\ref{eq:SCS_spinor}) leads to
\begin{equation}\label{eq:spinormf}
    H_{\rm S}=\frac{\Lambda}{2} (1-z^2) \cos^2{\varphi} + z + 1,
\end{equation}
by keeping the leading terms and omitting the constant ones, and once again $z= {\rm cos}\theta$ while $\Lambda = -2/q$. 
Note, the values of $\Lambda$ can be both negative and positive depending on the value of $q$. The negative value of $\Lambda$ does not change the structure of the mean-field phase space as discussed in Fig.~\ref{fig:phase_portrait_spinor}.

The three different regimes are also present in the case of the symmetric (anti-symmetric) subspace of the spinor system. To find positions of fixed points one should start with Hamilton equations for conjugate variables $(z, \varphi)$ using (\ref{eq:spinormf}), they are $\dot{\varphi} = - \Lambda z \cos^2\varphi +1 =0$, $\dot{z}= 2\Lambda(1 - z^2)\cos\varphi\sin\varphi =0$. 
Next, one calculates solutions of $(\dot{z}, \dot{\varphi})=(0,0)$ which are locations of fixed points.
The three regimes can be distinguished and they are listed below for negative values of $\Lambda$. The ``Rabi'' regime is in the limit $\Lambda\to 0$ when the evolution is governed by the linear term in the Hamiltonian. There are two stable center fixed points located at both poles of the Bloch sphere, i.e. $z = \pm 1$. It is true up to the bifurcation which occurs at $\Lambda=1$. On the other hand, in the ``Josephson'' regime, just after bifurcation, the fixed point at $z=-1$ became unstable and the two new stable center fixed points appear at $(z, \varphi)=(1/\Lambda , 0)$ and $(z, \varphi)=(1/\Lambda , \pi)$. In addition, the ``Fock'' regime takes place when the interaction term dominates over the linear one. This regime is characterized by the two stable center fixed points at $(z, \varphi)=(0, \pi/2)$ and $(z, \varphi)=(0, 3\pi/2)$ and the unstable along a meridian of the Bloch sphere at $\varphi=0, \, \pi$.

In our work we focus on the Josephson regime for $|\Lambda| =2$. 
The desired ``$\infty$'' shape is draw up by trajectories centered around an unstable fixed point. Moreover, the angle among constant energy lines incoming and outgoing from the saddle fixed point equals to $\pi/2$, see Figs.~\ref{fig:phase_portrait_bimodal} and  \ref{fig:phase_portrait_spinor}. It means that the level of entanglement generated is the largest and the fastest, see~\cite{Sorelli_2019} and Fig.~\ref{fig:scalling}(c). The phase portrait consists of one unstable and three stable fixed points among which two are symmetrically located around the unstable one. These two stable center fixed points serve to our protocol as we will use them to stabilize entanglement dynamics by locating the state around them.

\section{Twist-and-store Protocol}\label{sec:protocol}

\begin{figure}[hbt!]
5		
%
%

	\includegraphics[width=1\linewidth]{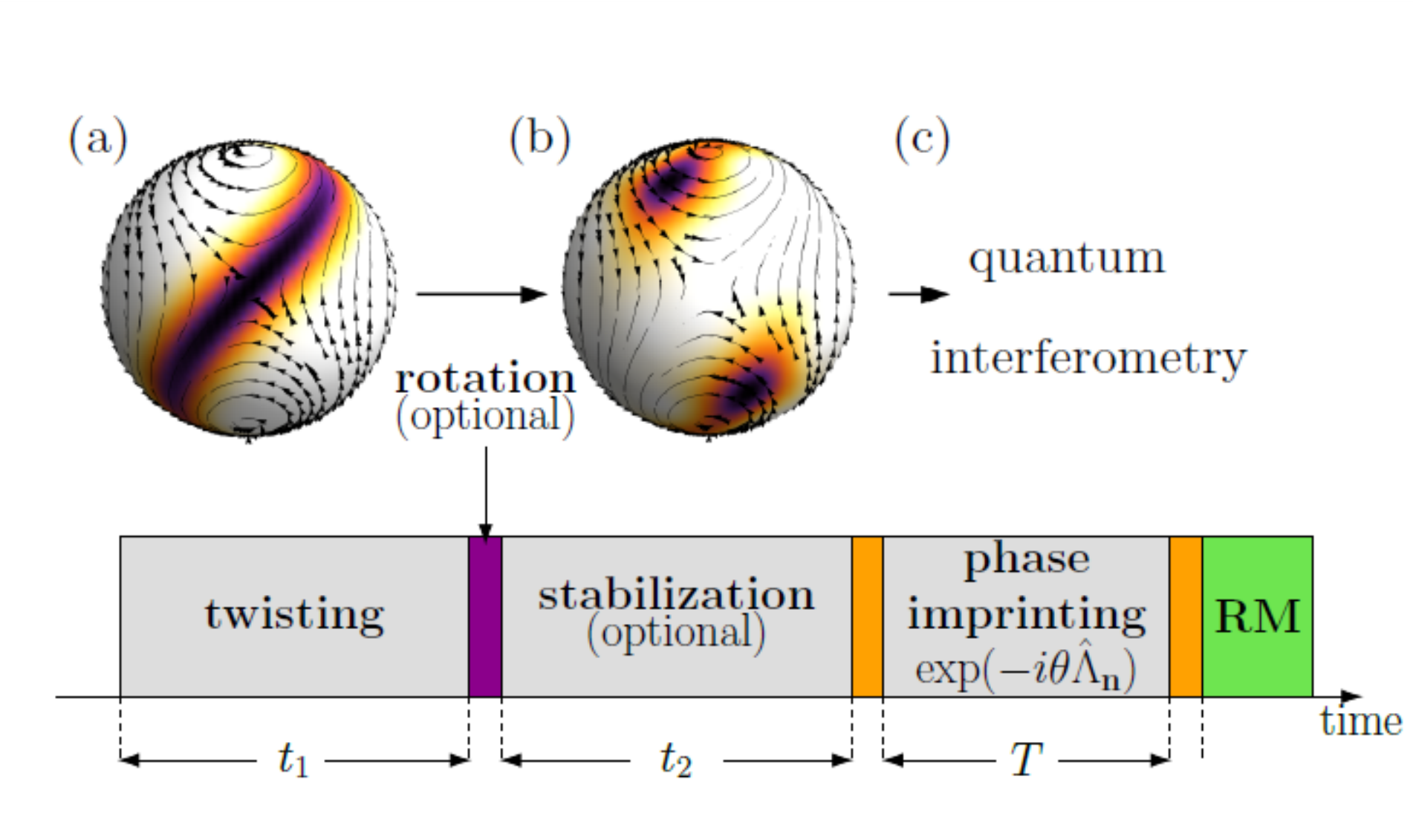}
	\caption{A protocol for entanglement storage for linear entanglement-enhanced quantum interferometry. (a) An entangled state is dynamically prepared by unitary evolution. (b) At the given moment of time, the state is rotated to location around stable fixed points and stabilization due to unitary evolution takes place. (c) The unitary evolution is followed by the phase $\theta$ accumulation during an interrogation time $T$ under generalized generator of interferometric rotation~$e^{-i \theta \hat{\Lambda}_{\bf n}}$. Finally, a readout measurement (RM) is performed. }
	\label{fig:fig2}

\end{figure}

The interferometric protocol we consider consists of four steps in general, see Fig.~\ref{fig:fig2}. The scheme starts with the dynamical state preparation by the unitary evolution determined by the system Hamiltonian followed by the state rotation at a given moment of time. The unitary evolution continues and eventually leads to the stabilization of dynamics around the two stable fixed points located symmetrically around the unstable saddle fixed point. This state can further be used in quantum interferometry protocol, which consists of the phase $\theta$ accumulation during an interrogation time $T$ under the generalized generator $\hat{\Lambda}_{\bf n}$ of interferometric rotation~$e^{-i \theta \hat{\Lambda}_{\bf n}}$. In particular, this is the phase encoding step in which the unitary  transformation ~$e^{-i \theta \hat{\Lambda}_{\bf n}}$ describes our interferometer in the language of the quantum  mechanics. The phase $\theta$ depends on the physical parameter to be measured, e.g. a magnetic field, and we assume that it is imprinted onto the state in the most general way. At the end, a readout measurement (RM) is performed.

In this paper, we consider the system at zero temperature and therefor its unitary evolution is given by the $\hat{U}_{\rm BI}=e^{-i t \hat{H}_{\rm BI}}$ operator for the bimodal and by $\hat{U}_{\rm S}=e^{-i t \hat{H}_{\rm S}}$ for the spin-1 systems.
The initial state is the spin coherent state located around the unstable saddle fixed point, $|\psi(0)\rangle_{\rm BI} = |0, \pi/2 \rangle_{\rm BI}$ for the bimodal system and $|\psi(0)\rangle_{\rm S} = |0, \pi \rangle_{\rm S}$ for the spin-1 system. Note, in the latter case the state is located on the south pole of the symmetric Bloch sphere and it is the polar state $|0,N,0\rangle$. The corresponding Schr\"odinger equations 
are solved numerically in the Fock state basis where operators are represented by matrices and states are represented by vectors.

\section{Quantifying entanglement}

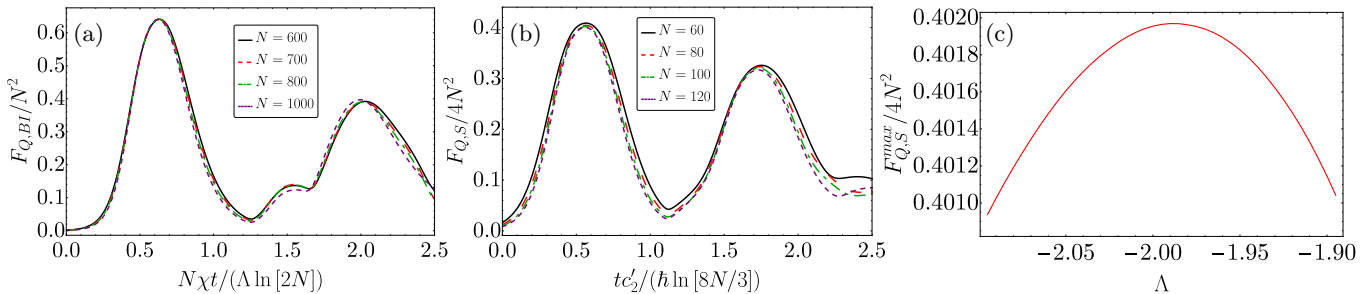
\begin{figure*}
    \centering
    \begin{picture}(240,120)
    \put(-115,95){(a)}
    \put(50,95){(b)}
    \put(230,95){(c)}
    \put(-140,0)
    {\includegraphics[width=1\linewidth]{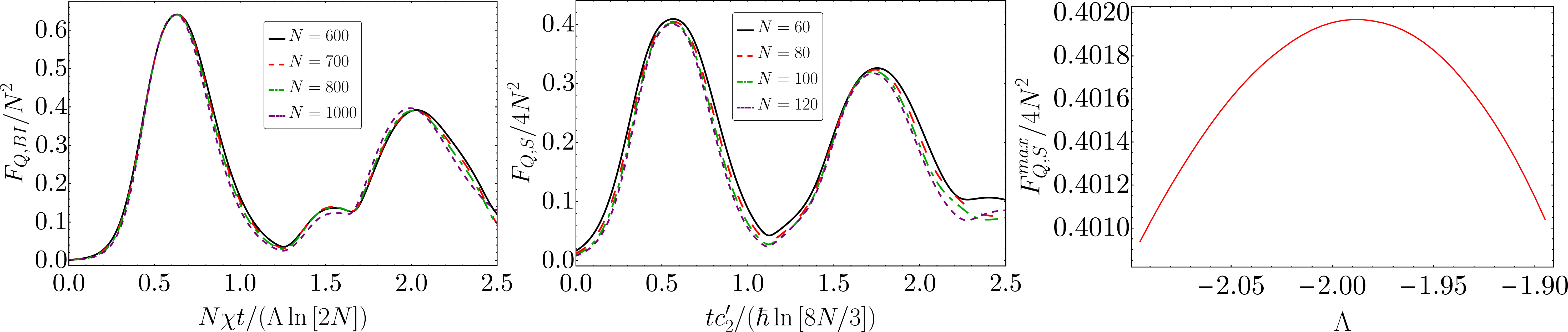}}
    \end{picture}
    \caption{The scaling of the quantum Fisher information with $N$ vs. time for the bimodal (a) and spinor (b) systems with $|\Lambda|=2$. The values of $N$ are given in the legend.
    (c) The maximal value of the QFI for spinor system, $F^{max}_{\rm Q,S}$, versus $\Lambda$ for $N=100$ demonstrating that the maximal value of entanglement is generated for  $\Lambda\simeq -2$.}
    \label{fig:scalling}
\end{figure*}

We measure the level of entanglement using the quantum Fisher information (QFI) because we consider the protocol in the context of quantum interferometry, as illustrated in Fig.~\ref{fig:fig2}.  It is already well established that the QFI is a good certification of entanglement useful for quantum interferometry~\cite{PhysRevA.85.022321}.

In a general linear quantum interferometer, the output state $|\psi(\theta)\rangle$ can be considered as the action of the rotation performed on the input state~$|\psi(t)\rangle$, namely $|\psi(\theta)\rangle = e^{-i\theta \hat{\Lambda}_{\mathbf{n}}}|\psi(t)\rangle$.
The QFI quantifies the minimal possible precision of estimating the imprinted phase $\theta$ in quantum interferometry~\cite{braunstein1994}. The minimal precision is given by the inverse of the quantum Fisher information $F_Q$, $\Delta \theta \geqslant 1/\sqrt{F_Q}$.
In general, the QFI value depends on the input state and generator of an interferometric rotation. The generator can be considered as the scalar product $\hat{\Lambda}_{\bf n}=\vec{\Lambda} \cdot {\bf n}$. The vector $\vec{\Lambda}$ is composed of bosonic Lie algebra generators describing a given system. Specifically, it is $\vec{{ \Lambda}}_{\rm BI}=\{ \hat{S}_x, \hat{S}_y, \hat{S}_y \}$ for bimodal and $\vec{{ \Lambda}}_{\rm S}=\{\hat{J}_x, \hat{Q}_{yz}, \hat{J}_y, \hat{Q}_{zx}, \hat{D}_{xy}, \hat{Q}_{xy},\hat{Y},\hat{J}_z\}$ for spinor condensates. The unit vector ${\bf n}$ determines the direction of rotation in the generalized Bloch sphere.

The QFI value is given by the variance
\begin{equation}\label{eq:fisher_information}
F_{Q} = 4 \Delta^2 \hat{\Lambda}_{\bf n},
\end{equation}
for pure states~\cite{Smerzi}. 
It is possible to find the generator $\hat{\Lambda}_{\bf n}$, for which the QFI reaches its maximum value~\cite{PhysRevA.82.012337}. For pure states, this problem can be solved by noticing that the variance in (\ref{eq:fisher_information}) can be written in terms of the covariance matrix
\begin{align}\label{eq:covariance_matrix}
\Gamma_{ij}[|\psi(t)\rangle]=\frac{1}{2} \langle \hat{\Lambda}_i \hat{\Lambda}_j + \hat{\Lambda}_j \hat{\Lambda}_i \rangle - \langle \hat{\Lambda}_i\rangle \langle \hat{\Lambda}_j \rangle ,
\end{align}
and then
\begin{equation}
    F_{Q} = 4\mathbf{n}^{T} \cdot \Gamma[|\psi(t)\rangle] \cdot \mathbf{n}.
\end{equation} 
Therefore, one concludes that the maximal value of the QFI is given by the largest eigenvalue $\lambda_{\rm max}$ of (\ref{eq:covariance_matrix}) while the direction of rotation ${\bf n}_{max}$ by the eigenvector corresponding to $\lambda_{\rm max}$.

There are two characteristic limits for the QFI value.
The first one is the standard quantum limit (SQL) typical for coherent states where the QFI is equal to $N$ for bimodal system and to $4 N$ for spinor system~\cite{Smerzi}. Whenever the QFI value is larger than the SQL, the state is entangled~\cite{PhysRevLett.96.010401}. The second is the Heisenberg limit which bounds the value of the QFI from above, and it is  equal to $N^2$ for bimodal system and $4 N^2$ for spinor system~\cite{Smerzi}. 

Here, we focus on the maximal value of (\ref{eq:fisher_information}) optimized over ${\bf n}$ at a given moment of time $t$ and the given input state $|\psi(t)\rangle_{\rm BI/S}$.
In the case of bimodal system, the maximal QFI is
\begin{equation}
    F_{Q,{\rm BI}}=4 \lambda_{\rm max, BI},
\end{equation}
where $\lambda_{\rm max, BI}$ is the maximal eigenvalue of the $3\times 3$ covariance matrix when $\hat{\Lambda}_i$ in (\ref{eq:covariance_matrix}) is replaced by $\hat{\Lambda}_{{\rm BI}, i}$. 
In Appendix \ref{app:covariancematrixspinor} we discuss the direction of interferometric rotation leading to the maximal value of the QFI.
In the case of spinor condensate, the QFI reads
\begin{equation}
    F_{Q,{\rm S}}=4 \lambda_{\rm max, S},
\end{equation}
where this time $\lambda_{\rm max, S}$ is the maximal eigenvalue of $8 \times 8$ matrix (\ref{eq:covariance_matrix}) when $\hat{\Lambda}_i$ is replaced by $\hat{\Lambda}_{{\rm S}, i}$. Although there are eight possible eigenvalues, only a few of them contribute to the maximal QFI value. It is because of the additional constant of motion, namely magnetization, which introduces symmetry of covariance matrix, simplifies its form and diminishes the number of various values of $\lambda_{S}$ and directions of interferometric rotations ${\bf n}$, see Appendix \ref{app:covariancematrixspinor} for details of calculations.

In Fig.~\ref{fig:scalling} we show an example of the QFI evolution in the Josephson regime for $|\Lambda|=2$ when $|\psi(t)\rangle_{\rm BI/S}=\hat{U}_{\rm BI/S} |\psi(0)\rangle_{\rm BI/S}$ (without optional rotation discussed in Fig.~\ref{fig:fig2} and in Section~\ref{sec:storage}). It was shown for bimodal condensates that for $|\Lambda|=2$ the unitary evolution generates the fastest speed and amount of entanglement~\cite{Sorelli_2019}. 
This is because of the characteristic ``$\infty$'' shape in the mean-field phase portrait with the angle between in- and out-going constant energy lines equals to $\pi/2$~\cite{Sorelli_2019, PhysRevA.92.013623}.
It is expected that this also holds true for spinor condensate due to the same characteristic shape drawn up by constant energy lines in the mean-field phase space.

It is interesting to note that the short time dynamics of the QFI exhibit a scaling behavior for a different number of particles, provided that the time axis is properly re-scaled as $Nt/{\rm ln}(2 N)$ for bimodal system and as $t/{\rm ln}(8 N/3)$ for spinor system (the difference in $N$ comes from the energy unit chosen for both systems). 
This can be interpreted as the appearance of the first maximum of $F_Q$ with Heisenberg scaling at $t \simeq \frac{\ln(2 N)}{N}$ ($t \simeq \ln(8 N/3) $) for the bimodal (spinor) condensate.
The scaling is demonstrated in Fig.~\ref{fig:scalling}. 
Indeed, curves corresponding to different number of atoms overlap for both bimodal and spinor systems.
The scaling can be explained using a theory developed in \cite{PhysRevA.65.053819} under two approximations. The first is the truncation of the Bogoliubov-Born-Green-Kirkwood-Yvon (BBGKY) hierarchy of equations of motion for expectation values of spin operators' products. We truncate the hierarchy by keeping the first- and the second-order moments, which is equivalent to the Gaussian approximation. The second approximation is the short-time expansion. The details of calculations are presented in Appendix \ref{app:scalingbimodal} for the bimodal and in Appendix \ref{app:scalingspinor} for the spinor systems. 

\section{Entanglement stabilization and storage around stable fixed points}\label{sec:storage}

The regular part of the initial evolution and structure of the mean-field phase space give a possibility of a stabilization scheme with nearly stationary value of the QFI at a relatively high level. The scheme consists of three steps, as discussed in Fig.~\ref{fig:fig2}. The first step is unitary evolution until the QFI reaches the value close to the first maximum.
Then, an instantaneous pulse rotates the state through $\alpha_{\rm BI}$ around the $\hat{S}_x$ axis,
\begin{equation}\label{eq:bomodalrot}
    |\psi(t_1^+)\rangle_{\rm BI}= e^{-i \alpha_{\rm BI} \hat{S}_{x}} |\psi(t_1^-)\rangle_{\rm BI}
\end{equation}
for the bimodal system, and through $\alpha_{\rm S}$ around the $\hat{J}_{zs}$ axis,
\begin{equation}\label{eq:spinorrot}
    |\psi(t_1^+)\rangle_{\rm S}= e^{-i \alpha_{\rm S} \hat{J}_{zs}} |\psi(t_1^-)\rangle_{\rm S}
\end{equation}
for the spin-1 system, where $t_1^{-}$ denotes the time just before  and $t_1^{+}$ after the rotation. Shortly before the rotation the Husimi function of the state is highly stretched. Rotation throws the  most stretched part of the state around stable regions of the phase space. Later on, for $t>t_{1}^+$, the state dynamics  is governed by the unitary evolution without any manipulations. However, it is trapped around the two stable fixed points.

\begin{figure*}
	\centering
	\begin{picture}(240,240)
	\put(-113,220){(a)}
	\put(45,220){(b)}
	\put(220,220){(c)}
	\put(-140,120)
	{\includegraphics[width=1\linewidth]{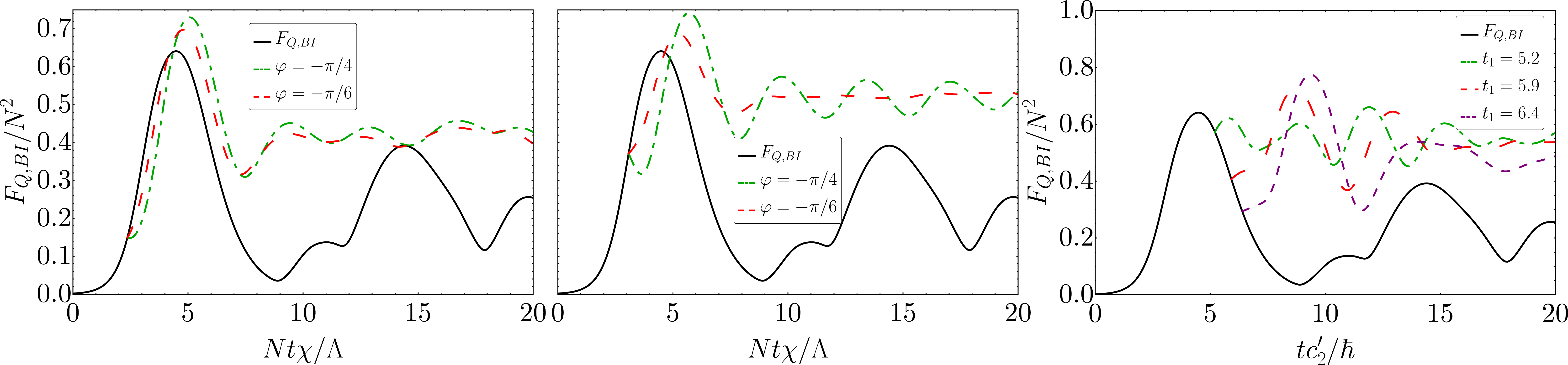}}
	\put(-113,90){(d)}
	\put(45,90){(e)}
	\put(220,90){(f)}
	\put(-140,-10)
	{\includegraphics[width=1\linewidth]{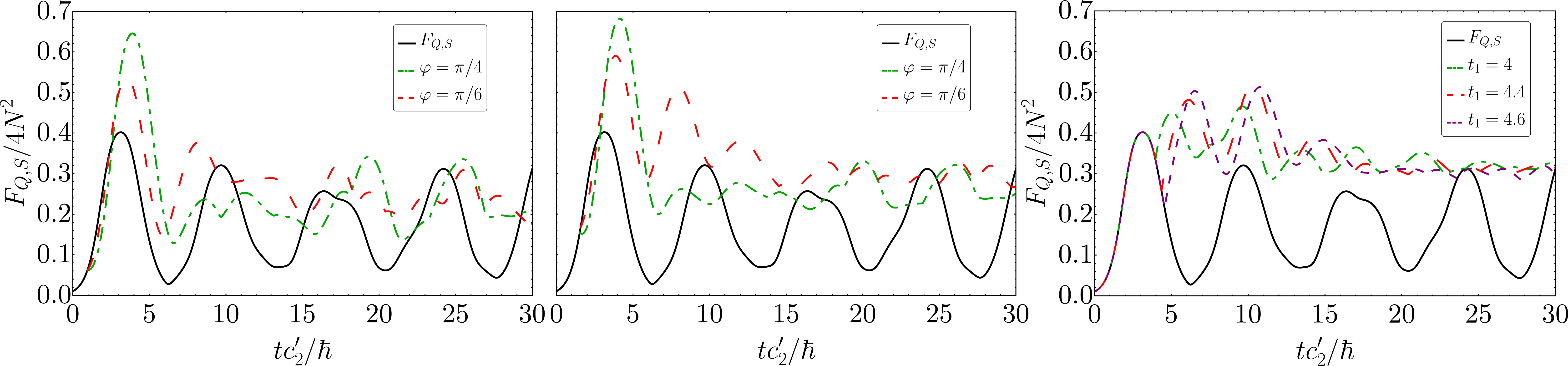}}
	\end{picture}
	\caption{The figures show QFI without (black lines) and with optional rotation at $t_1=2.37$ (a),  $3.08$ (b) for bimodal and $t_1=1$ (d), $1.6$ (e) spinor systems for $N=600$ and $N=100$, respectively. The two different values of rotation angle are considered $\alpha_{\rm BI/S}=\pi/4$ (green dash-dotted lines) and $\alpha_{\rm BI/S}=\pi/6$ (red dashed lines). Initially, for spin squeezed and a bit over-squeezed states the rotation of $\pi/4$ gives higher value of the QFI. However, for later times when the tails of the state start to turn around the two fixed points, just before the QFI maximum, the optimal rotation angle changes and we observe that for angle of $\pi/6$ the QFI stabilizes quicker on higher values. (c) and (f) show the QFI without (black lines) and with optional rotation of $\pi/6$ (color lines) after maximum for $N=600$ and $N=100$ for bimodal and spinor systems, respectively. }
	\label{fig:qfirotationN=100}
\end{figure*}
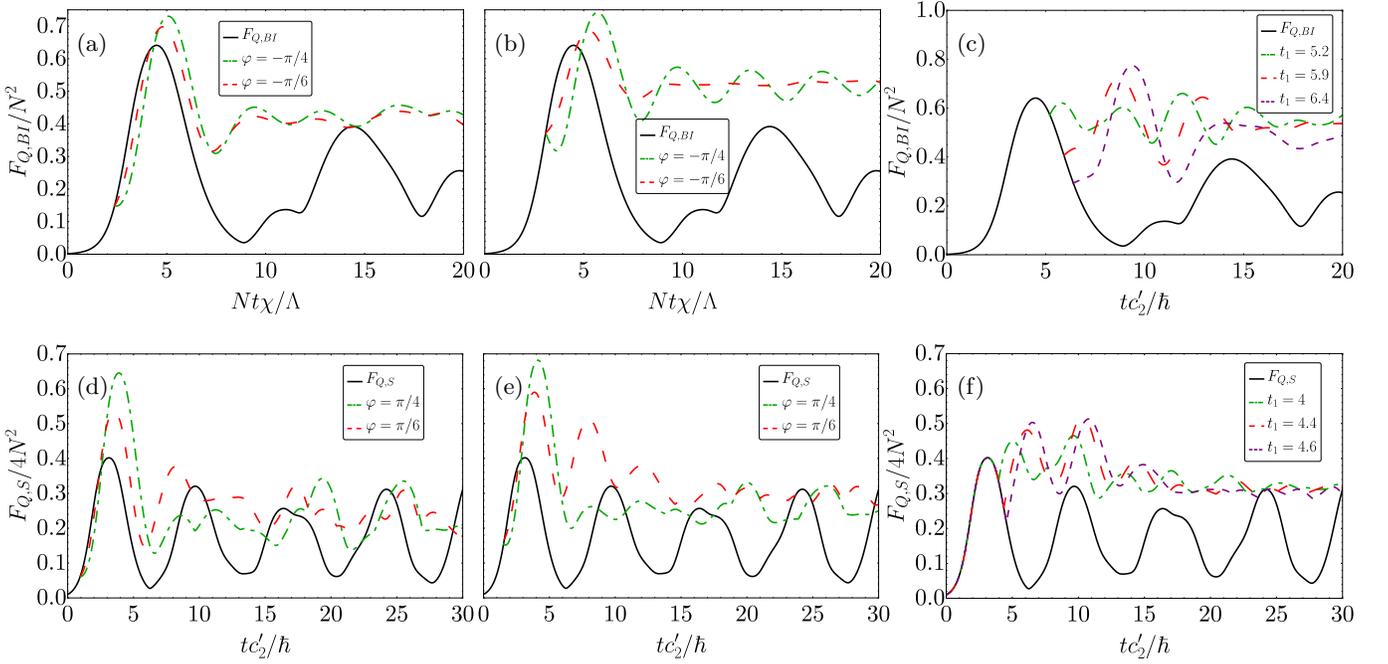

An example of the QFI is presented in Fig.~\ref{fig:qfirotationN=100}. 
An animation for time evolution of the Husimi function is shown in the Supplement Materials for the spinor system.
A roughly stationary value of the QFI is obtained in the long time limit. More interestingly, the twofold increase of the QFI value can be observed just after the rotation.
One might expect that the best rotation angle is $\pi/4$ as it is the intersection angle of the in- and out-going constant energy lines at the saddle fixed point.
It is true if the rotation takes place much before the QFI reaches its maximum, see Fig.~\ref{fig:qfirotationN=100}(a). At later times, a higher QFI value can be obtained for smaller values of the rotation angle, as demonstrated in Fig.~\ref{fig:qfirotationN=100}(c). This result does not depend much on the number of atoms while deviation from the optimal rotation time $t_1$ up to $20\%$ does not spoil the scheme, but rather lowers the QFI value.
Finally, we note that the rotation can also be performed after the QFI reaches the maximum. The slight increase of the QFI value is observed as well. This is illustrated in Fig.~\ref{fig:qfirotationN=100}.
All in all, we conclude, that it is advantageous to rotate the state in shorter times because of the fast gain in the QFI value.

It is intuitive that the QFI value stabilizes in the long time limit. When the state is located around the stable fixed point, the further dynamics are limited in this area of phase space and are approximately ''frozen". However, from the mathematical point of view it is non-trivial to show that indeed the value of the QFI, and therefore the entanglement, does not decrease in time. 
In below, we prove this for the bimodal condensate.

We assume that the direction of interferometric rotation just before the rotation is
$\hat{\Lambda}_{{\rm BI}, {\bf n}_{max}}(t_1^{-}) \approx \frac{1}{\sqrt{2}}(\hat{S}_z - \hat{S}_y)$ and therefore
$F_{Q,{\rm BI}}(t_1^{-})=4 \Delta^2 \left( \frac{\hat{S}_z - \hat{S}_y}{\sqrt{2}} \right)$, while after the rotation for $t\ge t_1^{+}$ one has
$\hat{\Lambda}_{{\rm BI}, {\bf n}_{max}}(t) \approx \hat{S}_z $ and  
$F_{Q,{\rm BI}}(t)=4 \Delta^2 \hat{S}_z$.
This is a fairly good approximation, as one can see in Appendix~\ref{app:covariancematrixbimodal}.
The QFI after rotation can be also written as
\begin{equation}
    F_{Q,{\rm BI}}(t)=\frac{4}{\hbar \chi} \left[ \langle \hat{H}_{\rm BI}(t) \rangle + 
    \hbar \Omega \langle \hat{S}_x(t)  \rangle \right],
\end{equation}
where we used the relation $\hbar \chi \langle \hat{S}_z^2\rangle = \hat{H}_{\rm BI} + \hbar \Omega \hat{S}_x$ employing (\ref{eq:bimodalH}) and $\langle \hat{S}_z \rangle =0$. Next, we note that an average energy is conserved after rotation, $\langle \hat{H}_{\rm BI}(t) \rangle = \langle \hat{H}_{\rm BI}(t_1^{+})\rangle$, while an average value of the $\hat{S}_x$ operator is bounded from below and above, namely $\frac{N}{2}\ge \langle \hat{S}_x \rangle \ge -\frac{N}{2}$. This two properties lead to the inequality
\begin{equation}\label{eq:fq_explanation}
    F_{Q,{\rm BI}}(t) \ge \frac{4}{\hbar \chi} \left[ \langle \hat{H}_{\rm BI}(t_1^{+}) \rangle - 
    \hbar \Omega \frac{N}{2} \right].
\end{equation}
The energy of the bimodal system after the rotation (\ref{eq:bomodalrot}) reads
$\langle \hat{H}_{\rm BI} (t_1^{+})\rangle = 
\hbar \chi \langle (\hat{S}_z(t_1^-) {\rm cos}\alpha_{\rm BI} + \hat{S}_y(t_1^-) {\rm sin}\alpha_{\rm BI} )^2\rangle - \hbar \Omega \langle \hat{S}_x(t_1^-) \rangle $, 
and for $\alpha_{\rm BI}=\pi/4$, it equals to
$\langle \hat{H}_{\rm BI} (t_1^{+})\rangle = 
F_{Q,{\rm BI}}(t_1^-) - \hbar \Omega \langle \hat{S}_x(t_1^-) \rangle $. Finally, one considers the latter term in (\ref{eq:fq_explanation}) to show that
\begin{equation}
    F_{Q,{\rm BI}}(t) \ge F_{Q,{\rm BI}}(t^-_1),
\end{equation}
for $t\ge t_1^+$ as $\langle \hat{S}_x(t_1^-) \rangle \ge -\frac{N}{2}$ as well.

The same reasoning can be used to demonstrate $F_{Q,{\rm S}}(t) \ge F_{Q,{\rm S}}(t^-_1)$ for spinor system, and we provide the calculation in Appendix~\ref{app:forspinors}.

\section{The parity operator as an efficient readout measurement}\label{sec:parity}

The precision of estimation of the unknown phase $\theta$ can be estimated using the signal-to-noise ratio as
\begin{equation}
\delta \theta^2
=\frac{\Delta^2{\hat{\mathcal{S}}}}{|\partial_\theta\ave{\hat{\mathcal{S}}}|^2} 
\label{eq:error-propagation}
\end{equation}
with $\Delta^2{\hat{\mathcal{S}}}=\ave{\hat{\mathcal{S}}^2}-\ave{\hat{\mathcal{S}}}^2$ representing the variance of the signal $\hat{\mathcal{S}}$ of which an average value is to be measured. Generally speaking, the precision in the $\theta$ estimation fulfils
\begin{equation}\label{eq:inequalities}
    \frac{1}{\delta \theta^2}\le F_Q.
\end{equation}
As mentioned before, the QFI gives the highest possible precision on estimation of $\theta$, but its measurement requires extracting the whole state tomography~\cite{Pezze2016-dx}. On the other hand, the inverse of the signal-to-noise ratio gives the lowest precision while it needs measurement of the first and second moments of the observable $\hat{\mathcal{S}}$ which is a bonus from the experimental point of view. 
 
On the one hand, in general $\hat{\mathcal{S}}$ is unknown. On the other hand, in some cases it is known as for example the parity operator for the Greenberger-Horne-Zeilinger (GHZ) state~\cite{GHZFirst, Bell_no_ineq_1990} or $\hat{J}_z^2$ for the spinor system~\cite{Niezgoda_2018}. Instead, the nonlinear squeezing parameter was recently proposed~\cite{PhysRevLett.122.090503} to saturate the QFI value at short times for bimodal condensates. However, the measurement of nonlinear squeezing parameter is related to the measurements of higher order moments and correlations.

Here, we show that the inverse of signal-to-noise ratio with the parity operator in the place of $\hat{\mathcal{S}}$ in (\ref{eq:error-propagation}) when $\theta \to 0$ saturates the QFI value, for both the bimodal and spinor systems. The parity operator is a well-defined quantum mechanical observable, but, unlike other quantum observables, it has no classical counterpart. However, it was understood that its measurement would be useful in quantum metrology~\cite{doi:10.1080/00107514.2010.509995,Pezze2016-dx} in both the optical and atomic domains when using non-Gaussian quantum states. 
The measurement of parity remains an experimental challenge as it requires a resolution at the level of a single particle, although it has been partially demonstrated experimentally~\cite{Song574, Sun2014, PhysRevLett.117.060504, PhysRevLett.86.5870, Andersen2019, Gao2019, PhysRevLett.123.050401, PhysRevX.10.011046}.

Let us first concentrate on the bimodal system. The parity operator we consider $\hat{\mathcal{S}}_{\rm BI}=(-1)^{\hat{S}_x-N/2}$ commutes with the bimodal Hamiltonian, $\left[ \hat{\mathcal{S}}_{\rm BI}, \hat{H}_{\rm BI}\right]=0$, and also with the rotation operator (\ref{eq:bomodalrot}), $\left[ \hat{\mathcal{S}}_{\rm BI}, e^{-i \alpha_{\rm BI} \hat{S}_{x}} \right]=0$. 
When the initial state $|\psi(0) \rangle_{\rm BI}$ is the eigenstate of $\hat{\mathcal{S}}_{\rm BI}$,
we have $\hat{\mathcal{S}}_{\rm BI} |\psi(0) \rangle_{\rm BI} = |\psi(0) \rangle_{\rm BI}$, and consequently $\hat{\mathcal{S}}_{\rm BI} |\psi(t) \rangle_{\rm BI} = |\psi(t) \rangle_{\rm BI}$. Finally, it is easy to show the relation $\hat{\Lambda}_{\rm BI, {\bf n}_{max}} \hat{\mathcal{S}}_{\rm BI} = - \hat{\mathcal{S}}_{\rm BI} \hat{\Lambda}_{\rm BI, {\bf n}_{max}}$ even if one considers a general form of the generator of interferometric rotation $\hat{\Lambda}_{\rm BI, {\bf n}_{max}}= a \hat{S}_z + b\hat{S}_y$ with any $a^2+b^2 = 1$, see Appendix~\ref{app:covariancematrixbimodal}. 

We use all the above-mentioned properties of the state and parity operator to calculate (\ref{eq:error-propagation}). To do this we expand an average value of the parity operator up to the leading terms in $\theta$, obtaining
${}_{\rm BI}\langle \psi(\theta) | \hat{\mathcal{S}}_{\rm BI} |\psi(\theta) \rangle_{\rm BI} = 1- 2 \theta^2 {}_{\rm BI}\langle \psi(t) | \hat{\Lambda}_{\rm BI, {\bf n}_{max}}^2 |\psi(t) \rangle_{\rm BI} + 0(\theta^3)$. Having that, the variance in (\ref{eq:error-propagation}) can be expressed as
\begin{equation}\label{eq:varBI}
    \Delta^2{\hat{\mathcal{S}}_{\rm BI}} = 4 \theta^2 {}\langle \hat{\Lambda}_{\rm BI, {\bf n}_{max}}^2 \rangle + 0(\theta^3),
\end{equation}
because $\langle \hat{\mathcal{S}}_{\rm BI}^2 \rangle =1$. The leading terms of the derivative in respect to $\theta$ of an average value of the parity is simply
\begin{equation}\label{eq:derBI}
    \partial_\theta  \langle \hat{\mathcal{S}}_{\rm BI}  \rangle 
    = -4 \theta  \langle \hat{\Lambda}_{\rm BI, {\bf n}_{max}}^2 \rangle + 0(\theta^2).
\end{equation}
Therefore, by inserting (\ref{eq:varBI}) and (\ref{eq:derBI}) into (\ref{eq:error-propagation}) it is possible to show that the leading terms in $\theta$ of the inverse of the signal-to-noise ratio  
\begin{equation}
\delta\theta^{-2} |_{\theta = 0} = 4 \Delta^2 \hat{\Lambda}_{\rm BI, {\bf n}_{max}},
\end{equation}
saturate the QFI value according to (\ref{eq:fisher_information}) due to the fact that $\ave{\hat{\Lambda}_{\rm BI, {\bf n}_{max}}}=0$. Note, the above derivation holds also with optional rotation of the state because the parity and rotation operators commute.

\begin{figure}
\begin{picture}(0,180)
    \put(-100,160){(a)}
    \put(15,160){(b)}
    \put(-100,75){(c)}
    \put(15,75){(d)}
    \put(-125,0)
    {\includegraphics[width=1\linewidth]{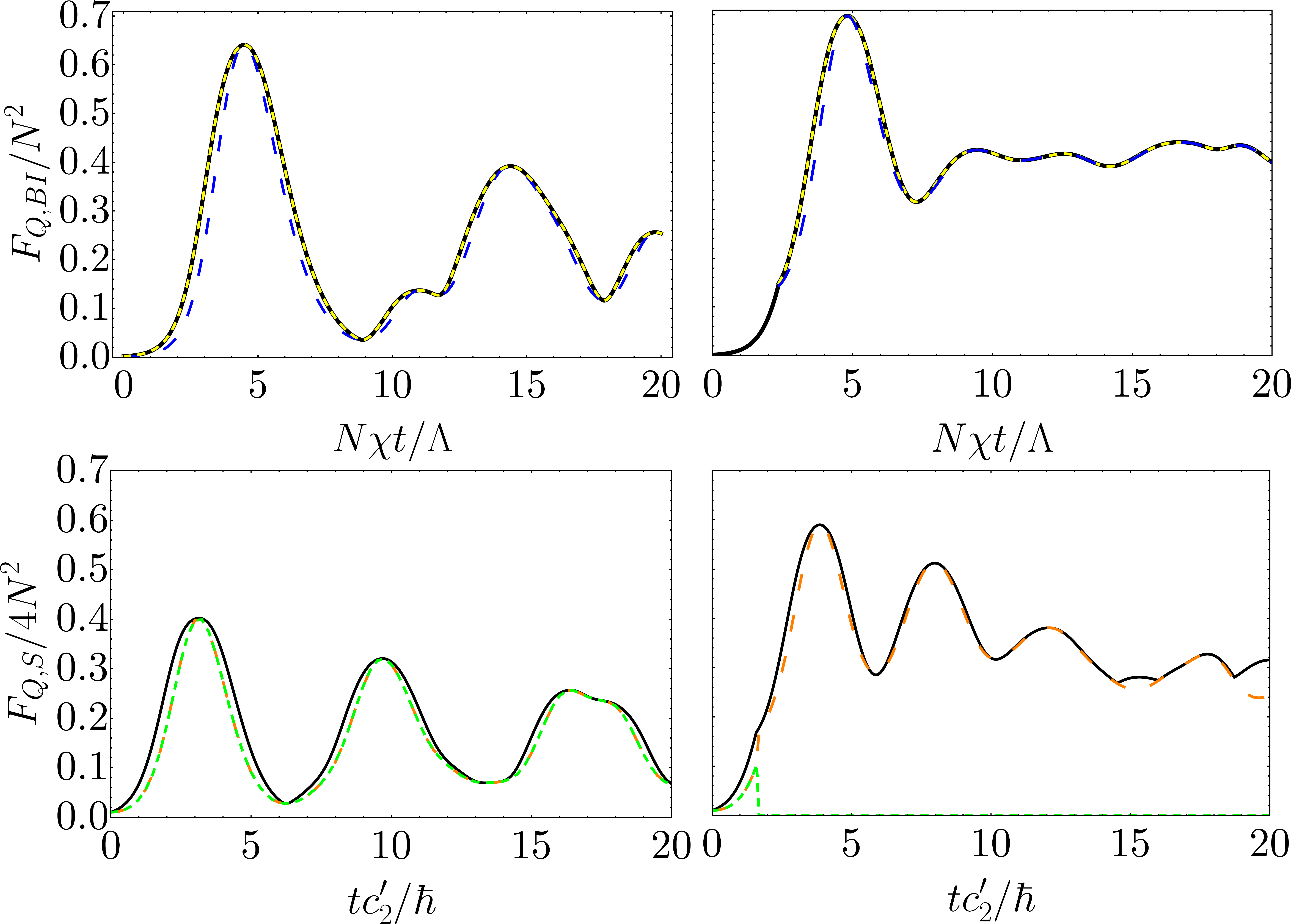}}
    \end{picture}
    \caption{(a-b) QFI (black solid) from the maximal eigenvalue of the covariance matrix (\ref{eq:covariance_matrix}) in bimodal system. 
    The error from the inverse of the signal to noise ratio when $\hat{\mathcal{S}}=(-1)^{\hat{S}_x-N/2}$ the parity measurement with the generator of interferometric rotation given by the eigenvector of the maximal eigenvalue of the covariance matrix (yellow dotted line) and when $\hat{\Lambda}_{\rm BI,{\bf n}_{max}} = \hat{S}_z$ (blue dashed line). Numerical results without rotation (a), and with rotation (\ref{eq:bomodalrot}) when $\alpha_{\rm BI}=\pi/6$ and $N \chi t_1/\Lambda= 2.36$ (b) for $N=600$.
    (c-d) QFI (black solid) from the maximal eigenvalue of the covariance matrix (\ref{eq:covariance_matrix}) in the spinor system for $N=100$. Here, $\hat{\Lambda}_{\rm S, {\bf n}_{max}} = \hat{J}_{y,a}$ and the inverse of the signal-to-noise ratio is shown by the orange dashed line in the case without optional rotation (c), and with the optional rotation (\ref{eq:spinorrot}) by $\alpha_{\rm S}$ at $t_1 c'_2/\hbar= 1.6$ (d). The green dotted line stands for $\hat{\mathcal{S}}= \hat{J}_z^2$ with the same $\hat{\Lambda}_{\rm S, {\bf n}_{max}} = \hat{J}_{y,a}$.}
    \label{fig:error}
\end{figure}
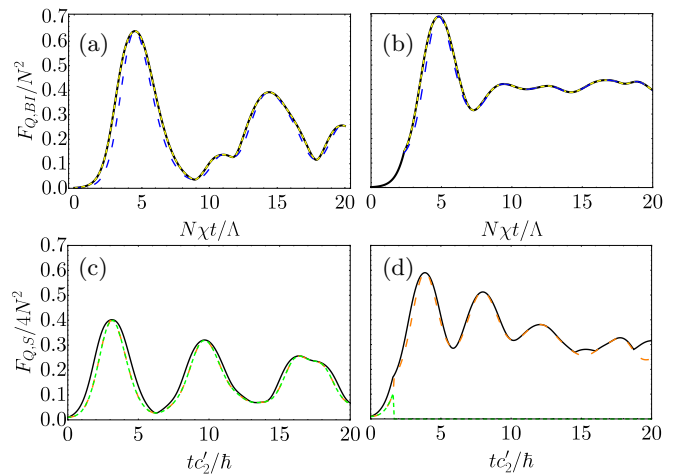

In Fig.~\ref{fig:error} we demonstrate our finding for the most optimal interferometer $\hat{\Lambda}_{\rm BI, {\bf n}_{max}}$ given numerically in Appendix~\ref{app:covariancematrixbimodal} (yellow dotted line) and simpler $\hat{\Lambda}_{\rm BI}=\hat{S}_z$ operator (blue dashed line) without (a) and with (b) optional rotation that locates the state around stable fixed points. The perfect agreement can be noticed.

Exactly the same reasoning can be applied for the spinor system but this time we define the parity as $\hat{\mathcal{S}}_{\rm S}=(-1)^{\hat{J}_{z,s}-N}$. One can show, by simple algebra, that parity commutes with the spinor Hamiltonian, $\left[ \hat{\mathcal{S}}_{\rm S}, \hat{H}_{\rm S}\right]=0$, and the optional rotation operator (\ref{eq:spinorrot}), $\left[ \hat{\mathcal{S}}_{\rm S}, e^{-i \alpha_{\rm S} \hat{J}_{z,s}} \right]=0$. 
The initial state is the eigenstate of the parity operator, $\hat{\mathcal{S}}_{\rm S} |\psi(0) \rangle_{\rm S} = |\psi(0) \rangle_{\rm S}$, 
and also any other state produced by the unitary evolution, 
$\hat{\mathcal{S}}_{\rm S} |\psi(t) \rangle_{\rm S} = |\psi(t) \rangle_{\rm S}$. 
A general form of the generator of interferometric rotation for the spinor system should be $\hat{\Lambda}_{\rm S, {\bf n}_{max}}= a \hat{J}_{x,s} + b\hat{J}_{y,s}$ with $a^2+b^2 = 1$, see Appendix~\ref{app:covariancematrixspinor}. One can follow the same calculations as for the bimodal system and consider the leading terms in $\theta$ of the inverse of the signal-to-noise ratio obtaining 
$\delta\theta^{-2} |_{\theta = 0} = 4 \Delta^2 \hat{\Lambda}_{\rm S, {\bf n}_{max}}$. The latter saturates the QFI value according to (\ref{eq:fisher_information}) because $\ave{\hat{\Lambda}_{\rm BI, {\bf n}_{max}}}=0$. 
The derivation also holds true with optional rotation of the state (\ref{eq:spinorrot}) as the parity and rotation operators do commute.
We illustrate our finding in Fig.~\ref{fig:error} without (c) and with (d) the optional rotation that locates the state around stable fixed points using various interferometers. In addition, we also illustrates that the simple signal $\hat{\mathcal{S}}_{\rm S}=\hat{J}_z^2$ saturates the QFI value when the optional rotation is not taken into account (see dashed green lines in Fig.~\ref{fig:error} (c) and (d)). The latter readout measurement is effective because the variance of magnetization is a constant of motion. Therefore, one can use the same treatment as in the case of parity to see that indeed the inverse of signal-to-noise ratio with $\hat{J}_z^2$ in the place of $\hat{\mathcal{S}}$ in (\ref{eq:error-propagation}) saturates the QFI value.

Finally, note that the sensitivity from the inverse of signal-to-noise ratio might be resistance against phase noise. This is the case when the operator describing the phase noise does commute with the parity operator. Then, the sensitivity from (\ref{eq:error-propagation}) does not change even for a convex mixture of quantum states, see calculation in~\cite{PhysRevA.93.022331}. This fact is not in contradiction with the convexity of the QFI~\cite{1054175}, which states that a convex mixture of quantum states contains fewer quantum correlations than the ensemble average.

\section{Discussion and conclusion}

In this work we have investigated theoretically the possibility of the entanglement stabilization in bimodal and spin-1 condensates. Our method utilizes the structure of the mean-field phase space. In particular, twisting dynamics of the spin coherent state initiated around an unstable saddle fixed point is enriched by a single rotation which locates the state around stable center fixed points. This allows for the generation of non-Gaussian states with the stable value of the QFI which exhibits Heisenberg scaling with a pre-factor of the order of one. We analyzed the method numerically and analytically proving $(i)$ the scaling of the QFI and time with total atoms number, $(ii)$ the lower bound of the QFI after optional rotation and $(iii)$ the optimal parity enhanced readout measurement.

In this paper, we have ignored the effects arising from any source of decoherence, such as a dissipative interaction with a heat reservoir or atomic losses. The decoherence effects will degrade the sensitivity in the $\theta$ estimation. If minimized, the entangled state stabilized by the scheme proposed here yields a higher resolution. However, we must stress that decoherence effects will degrade all schemes proposed to enhance interferometric measurements. Therefore, it might be necessary to make detailed comparisons of schemes with the incorporation of decoherence.

There is one other source of decoherence other than environmental, namely detection noise, which we would like to address in the context of the parity measurement. 
In the signal-to-noise ratio~(\ref{eq:error-propagation}), the effect of detection noise on moments of the operator $\hat{\mathcal{S}}$ in the large atoms number limit is the same as if it was replaced by $\hat{\tilde{\mathcal{S}}}=\hat{\mathcal{S}} + \hat{\delta}_\mathcal{S}$, where $\hat{\delta}_\mathcal{S}$ is an independent Gaussian operator satisfying $\ave{\hat{\delta}_\mathcal{S}}=0$ and $\ave{\hat{\delta}_\mathcal{S}^{2}}=\sigma^{2}$~\cite{PhysRevA.73.013814}. Therefore, it is clear that the detection resolution $\sigma^2 \lesssim 1$ is required to keep high sensitivity. Recent experiments with cold atoms demonstrated that the single atom imaging resolution has been achieved in the context of single trapped atoms and optical lattices using fluorescence imaging \cite{singleA,sherson2010single}, and also in the context of mesoscopic ensembles in a cavity, where the number of atoms is determined from the shift in the cavity frequency \cite{zhang2012collective}. More recently, near single atom resolution has been achieved in bimodal and spinor systems \cite{PhysRevLett.111.253001,qu2020probing} with the prospect of having higher resolution. That should be enough for a proof-of-principle demonstration of proposed by us measurement scheme.

The scheme we propose demonstrates yet another possibility for enhancement and storage of entanglement making use of the abstract nature of the mean-field phase space without turning-off interaction among atoms. Moreover, the inter-atomic interaction is desirable for the entanglement stabilization and storage. We argued possibility of the scheme resistance against phase noise. However, due to the non-ideal structure of the states stored they might lead to robust interferometric application~\cite{PhysRevX.6.041044}, which provides an interesting direction for a further work.

\section*{ACKNOWLEDGMENTS}

We thank K. Paw\l{}owski, A. Smerzi and P. Treutlein for discussion. EW and SM are supported by the Polish National Science Center Grants DEC-2015/18/E/ST2/00760. AN is supported by Project no. 2017/25/Z/ST2/03039, funded by the National Science Centre, Poland, under the QuantERA programme.

\appendix

\section{Spin-1 operators}\label{app:lambda}

 \begin{align}
 \hat{J}_{x} &\ =\ \frac{1}{\sqrt{2}}\left( \hat{a}^{\dag}_{\scriptscriptstyle{-1}}\hat{a}_{\scriptscriptstyle{0}} + \hat{a}^{\dag}_{\scriptscriptstyle{0}}\hat{a}_{\scriptscriptstyle{-1}} + \hat{a}^{\dag}_{\scriptscriptstyle{0}}\hat{a}_{\scriptscriptstyle{+1}} + \hat{a}^{\dag}_{\scriptscriptstyle{+1}}\hat{a}_{\scriptscriptstyle{0}}\right), \\
 \hat{Q}_{zx} &\ =\ \frac{1}{\sqrt{2}}\left( -\hat{a}^{\dag}_{\scriptscriptstyle{-1}}\hat{a}_{\scriptscriptstyle{0}} - \hat{a}^{\dag}_{\scriptscriptstyle{0}}\hat{a}_{\scriptscriptstyle{-1}} + \hat{a}^{\dag}_{\scriptscriptstyle{0}}\hat{a}_{\scriptscriptstyle{+1}} + \hat{a}^{\dag}_{\scriptscriptstyle{+1}}\hat{a}_{\scriptscriptstyle{0}}\right), \\
 \hat{J}_{y} &\ =\ \frac{i}{\sqrt{2}}\left( \hat{a}^{\dag}_{\scriptscriptstyle{-1}}\hat{a}_{\scriptscriptstyle{0}} - \hat{a}^{\dag}_{\scriptscriptstyle{0}}\hat{a}_{\scriptscriptstyle{-1}} + \hat{a}^{\dag}_{\scriptscriptstyle{0}}\hat{a}_{\scriptscriptstyle{+1}} - \hat{a}^{\dag}_{\scriptscriptstyle{+1}}\hat{a}_{\scriptscriptstyle{0}}\right), \\
 \hat{Q}_{yz} &\ =\ \frac{i}{\sqrt{2}}\left( -\hat{a}^{\dag}_{\scriptscriptstyle{-1}}\hat{a}_{\scriptscriptstyle{0}} + \hat{a}^{\dag}_{\scriptscriptstyle{0}}\hat{a}_{\scriptscriptstyle{-1}} + \hat{a}^{\dag}_{\scriptscriptstyle{0}}\hat{a}_{\scriptscriptstyle{+1}} - \hat{a}^{\dag}_{\scriptscriptstyle{+1}}\hat{a}_{\scriptscriptstyle{0}}\right), \\
 \hat{D}_{xy} &\ =\ \hat{a}^{\dag}_{\scriptscriptstyle{-1}}\hat{a}_{\scriptscriptstyle{+1}} + \hat{a}^{\dag}_{\scriptscriptstyle{+1}}\hat{a}_{\scriptscriptstyle{-1}}, \\
 \hat{Q}_{xy} &\ =\ i\left( \hat{a}^{\dag}_{\scriptscriptstyle{-1}}\hat{a}_{\scriptscriptstyle{+1}} - \hat{a}^{\dag}_{\scriptscriptstyle{+1}}\hat{a}_{\scriptscriptstyle{-1}}\right), \\
 \hat{Y} &\ =\ \frac{1}{\sqrt{3}}\left( \hat{a}^{\dag}_{\scriptscriptstyle{-1}}\hat{a}_{\scriptscriptstyle{-1}} - 2\hat{a}^{\dag}_{\scriptscriptstyle{0}}\hat{a}_{\scriptscriptstyle{0}} + \hat{a}^{\dag}_{\scriptscriptstyle{+1}}\hat{a}_{\scriptscriptstyle{+1}}\right),\\
 \hat{J}_{z} &\ =\  \hat{a}^{\dag}_{\scriptscriptstyle{+1}}\hat{a}_{\scriptscriptstyle{+1}} - \hat{a}^{\dag}_{\scriptscriptstyle{-1}}\hat{a}_{\scriptscriptstyle{-1}} ,
 \end{align}
 where $\hat{a}_{m_F}$ is the annihilation operator of the particle in the $m_F$ Zeeman component.

 \section{Anti-symmetric mean-field phase space}\label{app:meanfieldanti}
We will now address the equivalence of the anti-symmetric subspace. Similarly to the symmetric case, we are calculating an average value of (\ref{eq:spinorH}) over the spin coherent state defined for the anti-symmetric subspace as
\begin{equation}\label{eq:SCS_spinor_anti}
    |\varphi, \theta \rangle_{ a}=e^{-i\varphi \hat{J}_{z,a}/2} e^{-i\theta \hat{J}_{y,a}/2}|N0 \rangle_a
\end{equation}
where $|N0\rangle_a = \frac{\hat{g}_a^\dagger{}^N}{\sqrt{N!}} |000\rangle $ with $\varphi\in (0,2\pi),\, \theta\in (0,\pi)$. The spin coherent state (\ref{eq:SCS_spinor}) can be interpreted as a double rotation of the maximally polarized state $|N0\rangle_a$ in the anti-symmetric subspace and is an eigenstate of $\hat{J}_{z,a}$ with the eigenvalue $N$. Similarly to symmetric sphere, it is located on the north pole of the Bloch sphere.
In terms of spin-1 operators it reads $|N0\rangle_a = e^{-i \pi/4 \hat{Q}_{xy}}|00N\rangle$.
Just as in the symmetric case the state with N atoms in the $m_F=0$ mode is located on the south pole of the same Bloch sphere. 
To illustrate an arbitrary state $|\Psi \rangle $ on the Bloch sphere we use Husimi function $Q_a (\varphi, \theta)=|\langle \Psi | \varphi, \theta \rangle_a|^2$.
 
An average value of the spin-1 Hamiltonian (\ref{eq:spinorH}) over the spin coherent state (\ref{eq:SCS_spinor_anti}) leads to
\begin{equation}\label{eq:spinormf_anti}
    H_{a}=\frac{\Lambda}{2} (1-z^2) \sin^2{\varphi} + z - 1, 
\end{equation}
by keeping the leading and omitting the constant terms, and once again $z= {\rm cos}\theta$ while $\Lambda = -2/q$. Based on the difference between~\eqref{eq:spinormf_anti} and ~\eqref{eq:spinormf}, as well as the form of ~\eqref{eq:spinorHsymantysym}, one can see that phase portrait for anti-symmetrical subspace will be rotated by $\pi/2$ around $z-$axis.

\section{Structure of covariance matrix}\label{app:covariancematrixspinor}

It is interesting to find eigenvalues and eigenvectors for covariance matrix in the case of both systems. 
It is $3\times 3$ real matrix for the bimodal and $8\times 8$ real matrix for the spinor system, in general. However, in the latter case the structure of the matrix can be simplified significantly due to constrain of fixed magnetization by the evolution.
We distinguish here two cases of zero and non-zero fluctuation of the magnetization value.

\subsection{Bimodal system}\label{app:covariancematrixbimodal}

In the case of the bimodal system the optimal generator interferometric rotation can be found analytically when $\Omega=0$, see \cite{phdFerrini,Schulte_2020}. In the general case, the only analysis can be done numerically and, therefore, we present it  below. 

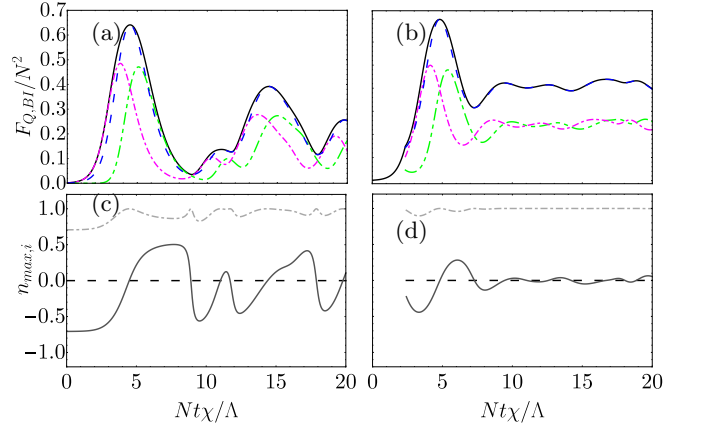
\begin{figure}[hbt!]
\begin{picture}(0,160)
    \put(-95,145){(a)}
    \put(20,145){(b)}
    \put(-95,80){(c)}
    \put(20,70){(d)}
    \put(-125,0)
    {\includegraphics[width=1\linewidth]{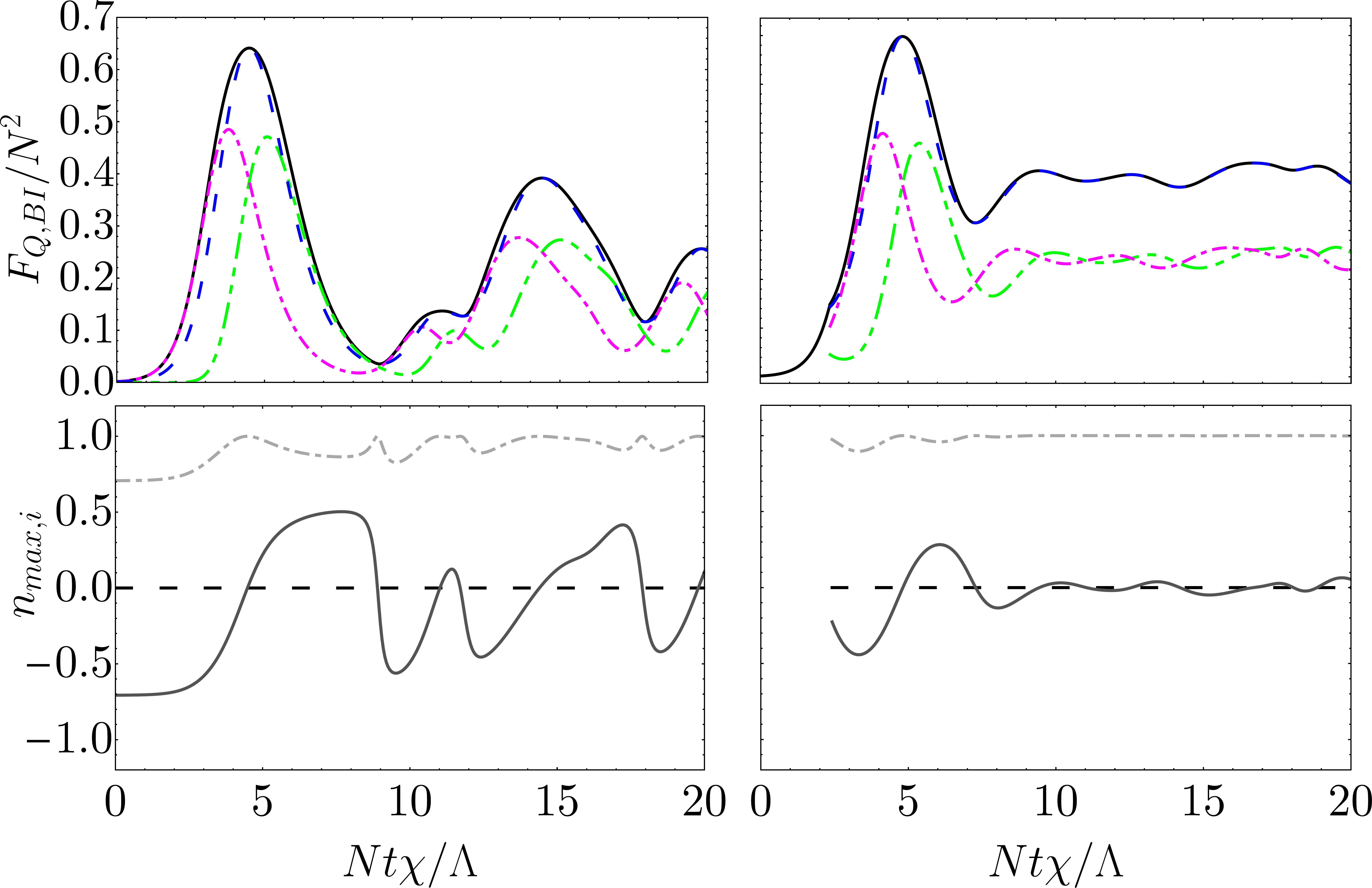}}
    \end{picture}
    \caption{(a-b) The QFI (black solid line) from the maximal eigenvalue of the covariance matrix (\ref{eq:covariance_matrix}), the QFI from (\ref{eq:fisher_information}) when $\hat{\Lambda}_{{\bf n}}=(\hat{S}_z - \hat{S}_y)/\sqrt{2}$ (purple dash-dotted line), $\hat{\Lambda}_{{\bf n}}=(\hat{S}_z + \hat{S}_y)/\sqrt{2}$ (green dash-double-dotted line) and $\hat{\Lambda}_{{\bf n}}=\hat{S}_z$ (blue dashed line). 
    (c-d) An illustration of the best direction of interferometric rotation ${\bf n}_{max}$, i.e. $i$-th component of the covariance matrix eigenvector (\ref{eq:covariance_matrix}) corresponding to the highest eigenvalue. The $x$ component is plotted with black dashed line, $y$ by the dark gray solid line and $z$ with the light gray dot-dashed line. 
    Left panels: without optional rotation of the state (\ref{eq:bomodalrot}). Right panels: when optional rotation is applied at $t_1 = 2.36$ with $\alpha_{BI}=-\pi/6$. The calculations are performed for $N=600$.
   }
    \label{fig:opt_interf_bi}
\end{figure}

In Fig.~\ref{fig:opt_interf_bi}(a) the black solid line shows the QFI (\ref{eq:fisher_information}) given by the maximal eigenvalue of the covariance matrix (\ref{eq:covariance_matrix}), and variances of various generators of interferometric rotation $\hat{\Lambda}_{\bf n}$ in direction ${\bf n}$ as indicated in the figure caption.  Indeed one can see that in the case without optional rotation, Fig.~\ref{fig:opt_interf_bi}(c), initially the generator of interferometric rotation is a superposition of $\hat{S}_y$ and $\hat{S}_z$ (purple dot-dashed line in (a)) which saturates the QFI value. On the other hand, we also observe that the variance of $\hat{S}_z$ estimates well overall variation of the QFI in time. When the optional rotation (\ref{eq:bomodalrot}) is applied, see Fig.~\ref{fig:opt_interf_bi}(b), the optimal rotation axis is also  given by $\hat{S}_z$ (dashed line). Therefore we conclude that the QFI is well estimated by $4 \Delta^2 \hat{S}_z$ while the optimal interferometric rotation is the $z$-axis of the Bloch sphere.

\subsection{Spinor system: fixed magnetization}

The Hamiltonian (\ref{eq:spinorH}) conserves the magnetization which means that 
we have $[\hat{H}, \hat{J}_z] = 0$. 
Thus the following occurs for an arbitrary state 
\begin{align}
    \vert\Psi_\varphi\rangle = e^{-i\varphi \hat{J}_z}e^{-i\hat{H}t}\vert\Psi\rangle 
    = e^{-i\hat{H}t}e^{-i\varphi M}\vert\Psi\rangle.
\end{align}
An action of the rotation operator $e^{-i\varphi \hat{J}_z}$ results in the phase factor given by the product of rotation angle and magnetization.
On the other hand, the QFI has the same value for both $\vert\Psi\rangle$ and $\vert\Psi_\varphi\rangle$, and therefore one has the condition
\begin{equation*}
 4\mathbf{n}^{T} \cdot \Gamma[\vert \Psi \rangle] \cdot \mathbf{n} =  4\mathbf{n}^{T} \cdot \Gamma[\vert \Psi_\varphi \rangle] \cdot \mathbf{n},
\end{equation*}
and so $\Gamma[\vert \Psi \rangle] = \Gamma[\vert \Psi_\varphi \rangle]$.
From the definition of covariance matrix \eqref{eq:covariance_matrix} one can see that 
\begin{align}
    \Gamma[\vert \Psi_\varphi \rangle]_{ij}= \langle \Psi \vert \frac{1}{2}(\hat{\Tilde{\Lambda}}_i\hat{\Tilde{\Lambda}}_j + \hat{\Tilde{\Lambda}}_j\hat{\Tilde{\Lambda}}_i) \vert \Psi \rangle - 
    \langle \Psi \vert \hat{\Tilde{\Lambda}}_i \vert \Psi \rangle\langle \Psi \vert \hat{\Tilde{\Lambda}}_j \vert \Psi \rangle,
\end{align}
where $\hat{\Tilde{\Lambda}}_i = e^{i\varphi \hat{J}_z}\hat{\Lambda}_i e^{-i\varphi \hat{J}_z}$. 
The rotation of the vector $\vec{\Lambda}_{\rm S}$ components gives:
\begin{subequations}
\begin{align}
    &e^{i\varphi \hat{J}_z} \hat{J}_x e^{-i\varphi \hat{J}_z} = \hat{J}_x \cos{\varphi} - \hat{J}_y\sin{\varphi} \\
    &e^{i\varphi \hat{J}_z}\hat{J}_y e^{-i\varphi \hat{J}_z} = \hat{J}_y \cos{\varphi} + \hat{J}_x\sin{\varphi} \\
    &e^{i\varphi \hat{J}_z} \hat{J}_z e^{-i\varphi \hat{J}_z} =\hat{J}_z \\
    &e^{i\varphi \hat{J}_z}\hat{Q}_{xy} e^{-i\varphi \hat{J}_z} = \hat{Q}_{xy} \cos{2\varphi} + \hat{D}_{xy}\sin{2\varphi} \\
    &e^{i\varphi \hat{J}_z}\hat{D}_{xy} e^{-i\varphi \hat{J}_z} = \hat{D}_{xy} \cos{2\varphi} - \hat{D}_{xy}\sin{2\varphi} \\
     &e^{i\varphi \hat{J}_z}\hat{Q}_{yz} e^{-i\varphi \hat{J}_z} = \hat{Q}_{yz} \cos{\varphi} + \hat{Q}_{zx}\sin{\varphi} \\
     &e^{i\varphi \hat{J}_z}\hat{Q}_{zx} e^{-i\varphi \hat{J}_z} = \hat{Q}_{zx} \cos{\varphi} - \hat{Q}_{yz}\sin{\varphi} \\
     &e^{i\varphi \hat{J}_z}\hat{Y} e^{-i\varphi \hat{J}_z} = \hat{Y}.
\end{align}
\end{subequations}
Therefore, one can distinguish the following groups of operators: $\{ \hat{J}_x,\hat{J}_y\}$, $\{ \hat{D}_{xy},\hat{Q}_{xy}\}$, $\{ \hat{Q}_{zx},\hat{Q}_{yz}\}$, $\{ \hat{J}_z\},\{\hat{Y}\}$, which rotations can be described with the operator:
$$
\hat{\mathcal{R}}_\varphi =  \left( \begin{array}{cc}
\cos{\phi} & -\sin{\phi}   \\
\sin{\phi}  &  \cos{\phi}  \end{array} \right).
$$
In fact, we can see that
\begin{equation}\label{eq:condition}
    \Gamma[\vert \Psi_\varphi \rangle] = M_\varphi \cdot\Gamma[\vert \Psi \rangle]\cdot M_\varphi^T,
\end{equation}
where the rotation matrix $M_\varphi$ is equal to\\
$
\left( \begin{array}{cccccccc}
\cos{\phi} & 0 & - \sin{\phi} & 0 & 0&0 &0 &0  \\
0 & \cos{\phi} & 0 &  \sin{\phi} & 0 & 0&0 &0 \\  
\sin{\phi} & 0 & \cos{\phi} & 0 & 0&0 &0 &0 \\
0 & -\cos{\phi} & 0 &  \sin{\phi} & 0 & 0&0 &0 \\  
0&0&0&0&\cos{2\phi}&-\sin{2\phi}&0&0\\
0&0&0&0&\sin{2\phi}&\cos{2\phi}&0&0\\
0&0&0&0&0&0&1&0\\
0&0&0&0&0&0&0&1
\end{array} \right).
$\\
From the relation \eqref{eq:condition} and $\Gamma = \Gamma^T$, one obtains a set of equations that determine the possible zero values of covariance matrix elements, for example:
$$
\begin{cases}
\Gamma_{11} \cos{\phi} + \Gamma_{13}\sin{\varphi} = \Gamma_{11} \cos{\phi} - \Gamma_{13}\sin{\varphi}, \\
\Gamma_{13} \cos{\phi} - \Gamma_{11}\sin{\varphi} = \Gamma_{13} \cos{\phi} - \Gamma_{33}\sin{\varphi},
\end{cases}
$$
which shows that $\Gamma_{33}=\Gamma_{11}$ and $\Gamma_{13}=0$. 
Solving all possible remaining equations will give conditions for all the elements of the covariance matrix, namely $\Gamma_{44}=\Gamma_{22}$, $\Gamma_{66}=\Gamma_{55}$, $\Gamma_{34}=-\Gamma_{12}$. Except for $\Gamma_{77}$, $\Gamma_{88}$ and elements listed in (\ref{eq: covariance}), all the remaining elements are zero. On the other hand $\Gamma_{88}$ is defined by variance of $\hat{J}_z$, which stands for fluctuations of magnetization, thus this element is 0 as well.
In the subspace of zero magnetization we arrive with the block diagonal structure of the covariance matrix:
\begin{equation}\label{eq: covariance}
\Gamma = \Gamma_{s} \oplus \Gamma_{a} \oplus [\Gamma_{55}] \oplus [\Gamma_{55}] \oplus [\Gamma_{77}] \oplus [0],
\end{equation}
where 
\begin{equation}
\Gamma_{s} = 
\left( 
	\begin{array}{cc}
    	\Gamma_{11} & \Gamma_{12} \\
        \Gamma_{12} & \Gamma_{22}
	\end{array}
\right),
\ \ \ 
\Gamma_{a} = 
\left( 
	\begin{array}{cc}
    	\Gamma_{11} & -\Gamma_{12} \\
        -\Gamma_{12} & \Gamma_{22}
	\end{array}
\right).
\end{equation}

\begin{figure}[hbt!]
    \includegraphics[width=1\linewidth]{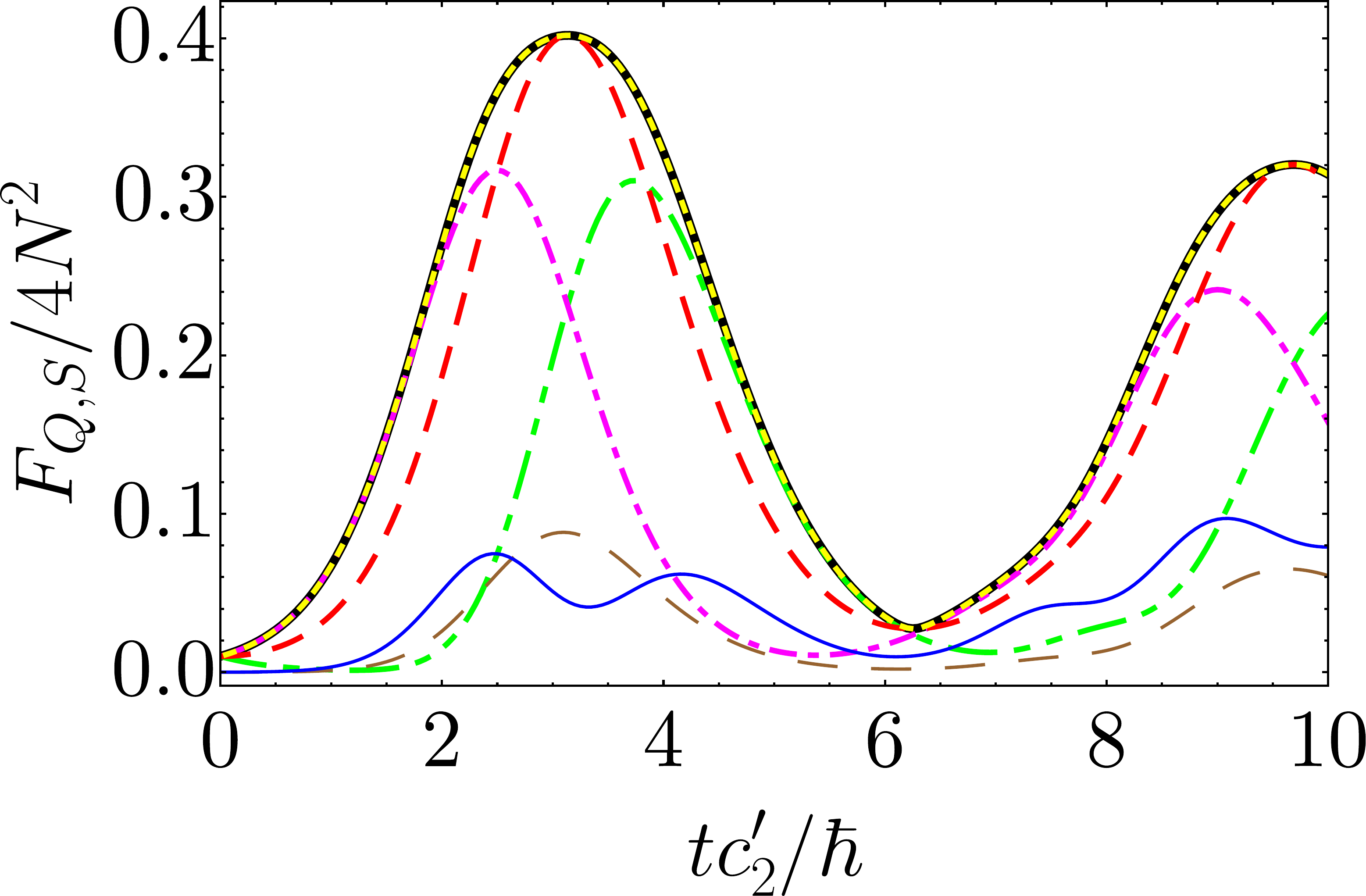}
    \caption{
    An illustration of optimal generators of interferometric rotation $\hat{\Lambda}_{{\rm S},{ij}}$ for spinor system with fixed magnetization given by (\ref{eqapp:C7})-(\ref{eqapp:C9}) calculated for $N = 100$ atoms. 
    The QFI optimalized over all directions ${\bf n}$ is shown by the black solid line.
    The corresponding values of the QFI for a given generator derived in the main text are:
    $F_{Q,{\rm S}}=4 \Delta^2 \hat{\Lambda}_{{\rm S},12}$ (which equals to $4 \hat{\Lambda}_{{\rm S},34}$) is marked by the yellow dashed line, 
    $F_{Q,{\rm S}}=4 \Delta^2 \hat{\Lambda}_{{\rm S},55}$ by the dashed brown thin line and $F_{Q,{\rm S}}=4 \Delta^2 \hat{\Lambda}_{{\rm S},77}$ by the thin blue line.
    The case with $\hat{\Lambda}_{{\rm S},ij} = \hat{J}_x$ is also shown for comparison by the dashed red line.
    In addition, the QFI with 
    $\hat{\Lambda}_{{\rm S},ij}=\frac{1}{\sqrt{2}}(\hat{J}_{xs} - \hat{J}_{ys})$ and $\hat{\Lambda}_{{\rm S},ij}=\frac{1}{\sqrt{2}}(\hat{J}_{xa} + \hat{J}_{ya})$ are shown by the purple dot-dashed and green dashed-double-dotted lines. The latter illustrates that the QFI value before the first maximum is given by $F_{Q,{\rm S}}=4 \Delta^2 \left( \frac{\hat{J}_{xs} - \hat{J}_{ys}}{\sqrt{2}}\right) $.}
    \label{fig:inter_norot-spinor}
\end{figure}

A diagonalization of the above matrix gives four possible generators of interferometric rotation~\cite{Niezgoda_2019}
\begin{align}
\hat{\Lambda}_{{\rm S},12}&  = \frac{\hat{\Lambda}_{1} - \gamma_{12}\hat{\Lambda}_{2}}{\sqrt{1 + \gamma_{12}^2}}, \label{eqapp:C7} \\
\hat{\Lambda}_{{\rm S},55}&  = \hat{\Lambda}_{5}, \label{eqapp:C8}\\
\hat{\Lambda}_{{\rm S},77}&  = \hat{\Lambda}_{7}, \label{eqapp:C9}
\end{align}
where $\gamma_{ij} =  (\Gamma_{jj}-\Gamma_{ii}-\sqrt{(\Gamma_{ii} -\Gamma_{jj})^2 + 4 \Gamma_{ij}})/(2 \Gamma_{ij})$. The corresponding values of the QFI are given by the variance
\begin{equation}
    F_{Q, {\rm S}}=4 \Delta^2 \hat{\Lambda}_{{\rm S},ij}.
\end{equation}
There are three possible values which depend on time.
It is worth noting that in the short times dynamics, it is $\hat{\Lambda}_{{\rm S},12}$ (or $\hat{\Lambda}_{{\rm S},34}$ as they are equivalent) that determines the QFI value. Moreover, we observe that it can be approximated by $\hat{\Lambda}_{1}$
without significant change in the QFI value, namely $\hat{\Lambda}_{{\rm S},12}\simeq \hat{\Lambda}_{1}=\hat{J}_x$. This is illustrated in Fig.~\ref{fig:inter_norot-spinor}.

\subsection{Spinor system: non-zero fluctuations of magnetization}

We consider here the more general case of the rotated state
\begin{align*}
    \vert\Psi_\varphi\rangle = e^{-i\varphi \hat{J}_{z,s}}e^{-i\hat{H}t}\vert\Psi\rangle,
\end{align*}
used by us in the main text to locate dynamics around stable fixed points. Here, $\hat{J}_{z,s}= \frac{1}{2}(\hat{D}_{xy} + \sqrt{3}\hat{Y})$. 
The analysis presented in the previous subsection is not valid because $[\hat{J}_{z,s}, \hat{H}]\neq 0$. Moreover, the state after rotation is no longer in the subspace of zero magnetization but it is spread over all subspaces of even magnetization. Therefore, it has non-zero fluctuations of magnetization.

To calculate elements of the covariance matrix we used Eq.~\eqref{eq:covariance_matrix}, 
where an average is taken over a general state $\vert k \rangle = \sum_{M,n}C_{M,n} \vert n, M+N-2n, n-M \rangle$ which coefficients of decomposition in the Fock state basis are $C_{M,n} \equiv C_{n, M+N-2n, n-M}$ resulting from the symmetry of rotation around $\hat{J}_{z,s}$.
The summation over $n$ depends on the sign of the $M$: from max$(0, M/2, M)$ to min$(M, \frac{N+M}{2}, M+N)$ while $-N<M<N$.
Due to the rotation, the system has non-zero variance of magnetization $\Delta \hat{J}_z$ which is constant in time.
In addition, the possible eigenvalues of $\hat{J}_z$ can only be even, i.e. $M \in \{-N, -N+2,...,N-2, N\}$ assuming $N$ is even as well due to symmetry of rotation operator $\hat{J}_{z,s}$. Therefore, $C_{M,n} = C_{-M,n-M}$. 

We can distinguish operators that change magnetization by $\pm 1$, they are $\{\hat{J}_x, \hat{Q}_{yz}, \hat{J}_y, \hat{Q}_{zx}\}$, by $\pm 2:\, \{ \hat{D}_{xy}, \hat{Q}_{xy}\}$ and by $0$: $\{\hat{Y},\hat{J}_z\}$. The mean value of operators from the group $\{\hat{J}_x, \hat{Q}_{yz}, \hat{J}_y, \hat{Q}_{zx}\}$ is zero since the state is spread over subspaces of even magnetization. Moreover, a mean value of product of operators that change magnetization by odd value are zero. We use this fact while calculating the covariance matrix elements $\Gamma_{ij}$: with subscript $i$ for the operator from the group $\{\hat{J}_x, \hat{Q}_{yz}, \hat{J}_y, \hat{Q}_{zx}\}$ and $j$ from$ \{\hat{D}_{xy}, \hat{Q}_{xy}, \hat{Y}, \hat{J}_{z}\}$. The second property that should be taken into account is the symmetry of the state, namely 
$C_{M,n} = C_{-M,n-M}$, which sets the elements like $\Gamma_{14}$ or $\Gamma_{58}$ to zero.

After careful consideration of all covariance matrix elements, one can show that it simplifies to
\begin{equation}\label{eq:covariance_rot}
\Gamma_{\rm S} = \Gamma_{s} \oplus \Gamma_{a} \oplus \Gamma_{r},
\end{equation}
for the spinor system, where 
\begin{equation*}
\Gamma_{s} = 
\left( 
	\begin{array}{cc}
    	\Gamma_{11} & \Gamma_{12} \\
        \Gamma_{12} & \Gamma_{22}
	\end{array}
\right),
\ \ \ 
\Gamma_{a} = 
\left( 
	\begin{array}{cc}
    	\Gamma_{33} & \Gamma_{34} \\
        \Gamma_{34} & \Gamma_{44}
	\end{array}
\right), 
\end{equation*}
\begin{equation*}
\centering
\Gamma_{r} = 
\left( 
	\begin{array}{cccc}
    	\Gamma_{55} & 0 & \Gamma_{57} & 0 \\
        0& \Gamma_{66} &0 & \Gamma_{68} \\
        \Gamma_{57} & 0 & \Gamma_{77} & 0 \\
        0 & \Gamma_{68} & 0 & \Gamma_{88} 
	\end{array}
\right).
\end{equation*}

\begin{figure}[hbt!]
    \includegraphics[width=1\linewidth]{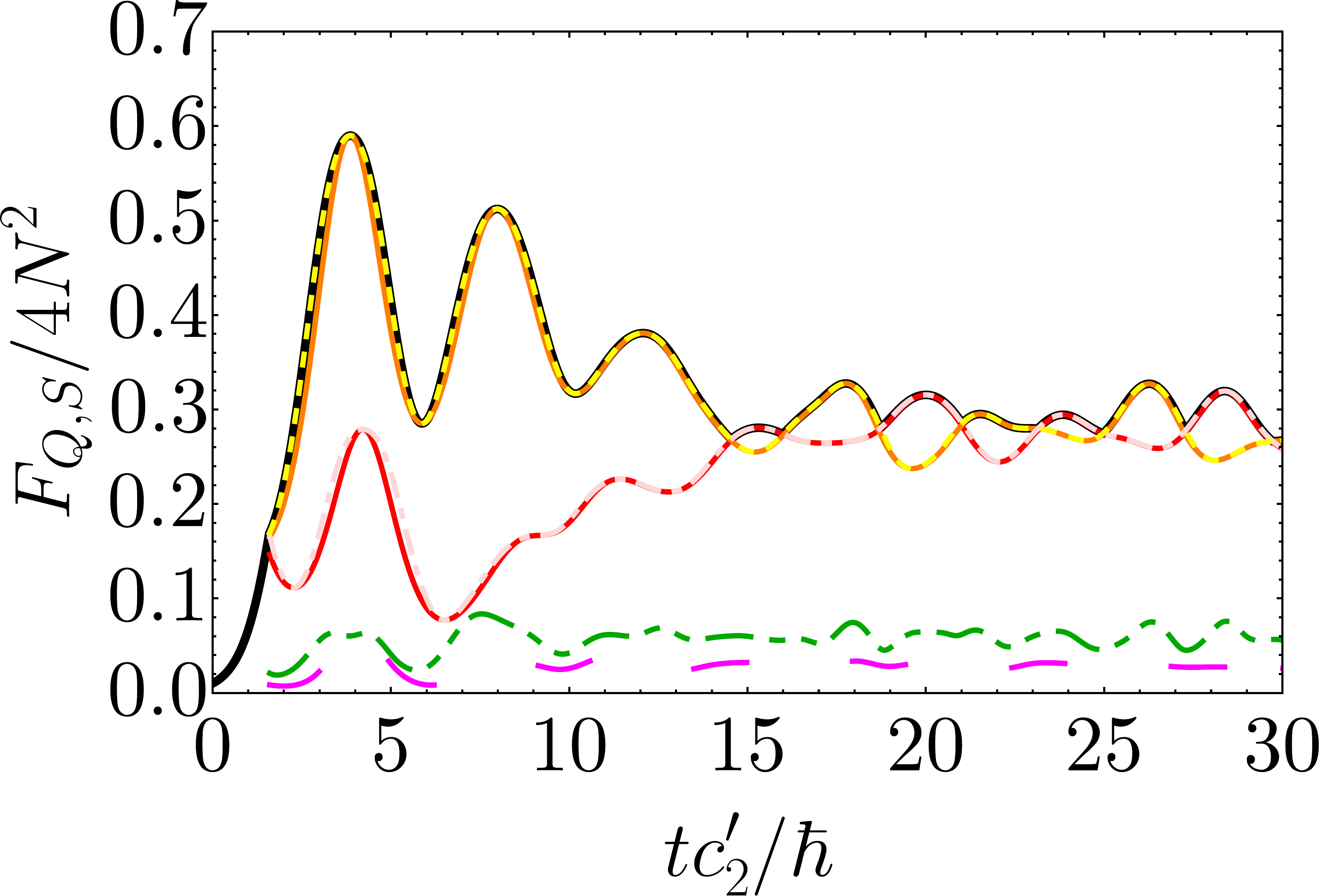}
    \caption{
    An illustration of optimal generators of interferometric rotation $\hat{\Lambda}_{{\rm S},{ij}}$ for spinor system with fluctuating magnetization given in (\ref{eqapp:C13}) calculated for $N = 100$ atoms. The relevant example discussed in the main text for states after the rotation around $\hat{J}_{zs}$ by $\pi/6$ at $t_1=1.6$. 
    The QFI optimized over all ${\bf n}$ is shown by the black solid line.
    The corresponding values of the QFI for particular generators are shown with $\hat{\Lambda}_{S,12}$ (light pink dash-dotted line), $\hat{\Lambda}_{S,34}$ (yellow dotted line), $\hat{\Lambda}_{S,57}$ (green dash-double-dotted line), $\hat{\Lambda}_{S,68} $ (purple dashed line). Finally, the QFI with 
    $\hat{\Lambda}_{{\rm S},ij}=\hat{J}_{xs}$ and $\hat{\Lambda}_{{\rm S},ij}=\hat{J}_{ya}$ are shown by the red and orange solid lines, respectively. The latter demonstrates that the QFI value after the rotation can be approximated well by $F_{Q,{\rm S}}=4 \Delta^2 \hat{J}_{ya}$.
    }
    \label{fig:interfer_rot-spinor}
\end{figure}

Diagonalization of \eqref{eq:covariance_rot} gives the following eigenvalues:
\begin{align}
    \lambda^{(\pm)}_{{\rm S},ij} = \frac{\Gamma_{{\rm S},ii}+\Gamma_{{\rm S},jj} \pm \sqrt{(\Gamma_{{\rm S},ii}-\Gamma_{{\rm S},jj})^2 + 4 \Gamma_{{\rm S},ij}^2}}{2},
\end{align}
where the pairs of indexes $(i,j)$ are one of $(1,2), (3,4), (5,7), (6,8)$. The contribution to the maximal value of the QFI can be from $\lambda^{(+)}_{{S},ij}$ for which the four possible generators of interferometric rotation are
\begin{equation}\label{eqapp:C13}
    \hat{\Lambda}_{{\rm S},ij}  = \frac{\hat{\Lambda}_{j} - \gamma_{ij}\hat{\Lambda}_{j}}{\sqrt{1 + \gamma_{ij}^2}}
\end{equation}
where $\gamma_{ij} =  (\Gamma_{jj}-\Gamma_{ii}-\sqrt{(\Gamma_{ii} -\Gamma_{jj})^2 + 4 \Gamma_{ij}})/(2 \Gamma_{ij})$. The corresponding values of the QFI determined by (\ref{eqapp:C13}), namely
\begin{equation}
F_{Q, \, {\rm S}}=4 \Delta^2 \hat{\Lambda}_{{\rm S},ij},
\end{equation}
are demonstrated in Fig.~\ref{fig:interfer_rot-spinor}.

\section{Scaling of the QFI for bimodal system}\label{app:scalingbimodal}

In order to analyze scaling of the QFI with the system size, we use a general theory developed in \cite{PhysRevA.65.053819}. One starts with equations of motion for operators of spin components which involve terms that depend on the first-order and second-order moments. Then, the time evolution of the second-order moments depends on second- and third-order moments, and so on. It leads to the Bogoliubov-Born-Green-Kirkwood-Yvon hierarchy of equations of motion for expectation values of operator products. We truncate the hierarchy by keeping the first- and the second-order moments.
\begin{align}
    \langle \hat{S}_i \hat{S}_j \hat{S}_k\rangle &\simeq
    \langle \hat{S}_i \hat{S}_j \rangle \langle \hat{S}_k\rangle +
    \langle \hat{S}_j \hat{S}_k \rangle \langle \hat{S}_i\rangle +
    \langle \hat{S}_k \hat{S}_i \rangle \langle \hat{S}_j\rangle \nonumber \\
    &- 2\langle \hat{S}_i \rangle \langle \hat{S}_j \rangle \langle \hat{S}_k\rangle.
\end{align}

Let us first rotate the Hamiltonian (\ref{eq:bimodalH}) around the $x$-axis of the Bloch sphere through $\pi/4$. The reason is as follows: there is nonzero angle between the constant energy line outgoing from the saddle fixed point and the $z$-axis. This angle is close to $\pi/4$ for $\Lambda=2$. Rotation of the Hamiltonian corresponds to the same rotation of the mean field phase portrait. It results in location of the constant energy line outgoing from the saddle fixed point along the $y$-axis, see Fig.~\ref{fig:phase_portrait_bimodal}. Note, the largest fluctuations that determine the QFI value are now located along the $y$-axis. Next, we introduce a small parameter $\varepsilon= 1/N$ and transform spin components into $\hat{h}_j=\sqrt{\varepsilon} \hat{S}_j$ while the commutation relations to $[\hat{h}_i,\hat{h}_j] =i\sqrt{\varepsilon} \hat{h}_k \epsilon_{ijk}$.
The rotated Hamiltonian (\ref{eq:bimodalH}) is
\begin{equation}
    \hat{H}=\frac{1}{\sqrt{\varepsilon}}
    \left( \hat{h}_z^2 + \hat{h}_y^2 
    + \hat{h}_z\hat{h}_y +  \hat{h}_y\hat{h}_z
    - a \hat{h}_x \right),
\end{equation}
where $a=2\varepsilon \Omega/\chi$, the energy unit is set to $\hbar \chi/(2\sqrt{\varepsilon})$ and we introduced dimensionless time $\tau = \chi t/(2 \sqrt{\varepsilon})$. 

Equations of motion for expectation values $s_j=\langle \hat{h}_j\rangle$ and second order moments $\delta_{jk}=\langle \hat{h}_j \hat{h}_k + \hat{h}_k \hat{h}_j\rangle  - 2 \langle \hat{h}_j \rangle \langle \hat{h}_k \rangle$ relevant for our purposes are
\begin{align}
    \dot{s}_x&= (\delta_{zz} - \delta_{yy}) \label{eq:sx},\\
    \dot{\delta}_{zz}&=-4 \delta_{zz}s_x-2a \delta_{yz},\label{eq:deltazz} \\
    \dot{\delta}_{yy}&=4 \delta_{yy}s_x+2a \delta_{yz}, \label{eq:deltayy}
\end{align}
The initial spin coherent state $|0, \pi/2 \rangle_{\rm BI}$ gives the following initial conditions: $s_x(0)=1/(2\sqrt{\varepsilon})$ and $\delta_{zz}(0)=\delta_{yy}(0)=1/2$. 

The equation (\ref{eq:deltazz}) is a non-homogeneous differential equation. The solution of its homogeneous part ($a\to 0$ in Eq.~(\ref{eq:deltazz})) is $\delta_{zz}(\tau)=\delta_{zz}(0)e^{-f(\tau)}$ with $f(\tau)=4 \int_0^\tau s_x(t)dt$. 
The analysis of non-homogeneous equation can be done by setting $\delta_{zz}(\tau)=C(\tau) e^{-f(\tau)}$ with $C(\tau)=C(0) - a \int_0^\tau \delta_{yz}(t) e^{f(t)}dt=\delta_{zz}(0)+\Phi(\tau)$. The part $\Phi(\tau)$ is very small and it can be omitted because of two reasons. Firstly, $\Phi(\tau)$ is of the order of small parameter $\varepsilon$. Secondly, in the short time expansion (up to the second order) one can indeed see that $\Phi(\tau)\simeq \Phi(0) + \dot{\Phi}(0) \tau =0$ due to $\delta_{yz}(0)=0$. Therefore, we conclude that the solution of (\ref{eq:deltazz}) can be well approximated by the solution of its homogeneous part.
The same analysis can be performed on Eq.~(\ref{eq:deltayy}) leading to $\delta_{yy}(\tau)=\delta_{yy}(0)e^{f(\tau)}$.
Eq.~(\ref{eq:sx}) takes the form $\dot{s}_x(\tau)=-  {\rm sinh}\left[ f(\tau) \right]$, that has an analytical solution when one expands the function $f(\tau)$ up to the first order in Taylor series $f(\tau) \simeq f(0) + \dot{f}(0)\tau$. The self-consistency condition gives $f(0) =0$ and $\dot{f}(0) = 4s_x(0)$. The approximated solution for $s_x$ takes the form~\cite{PhysRevA.92.013623}
\begin{equation}
    s_x(\tau)=s_x(0)-\frac{{\rm cosh}(4s_x(0)\tau)-1}{4s_x(0)},
\end{equation}
while the variance in the $y$ direction reads
\begin{equation}
\delta_{yy}=\delta_{yy}(0)e^{4s_x(0)\tau - \frac{{\rm sinh}(4 s_x(0)\tau) - 4 s_x(0)\tau}{[4 s_x(0)]^2}}.
\end{equation}

It can be shown by maximization of the QFI over the time resolves in the scaling of the first maximum as $\chi t_{max}\simeq {\rm ln}(2N)/N$. The leading term of the QFI maximum at the best time gives $F_{Q,{\rm BI}} \simeq 4 \Delta^2\hat{S}_y \simeq \frac{2}{\varepsilon} \delta_{yy}\simeq \frac{2}{e}\frac{1}{\varepsilon^2}\approx 0.7 N^2$.

\section{Scaling of the QFI for spinor system}\label{app:scalingspinor}

\begin{table*}[hbt!]
\centering
\caption{List of commutation relations among SU(3) algebra generators and spin components in the symmetric and anti-symmetric subspace used in this paper.}
\label{tab:tab1}
\begin{tabular}{||c|| c | c| c|c| c| c| c| c| c| c|} 
 \hline
$ {}_{T_i} {}^{T_j}$ & $\hat{J}_x (\hat{J}_{x,s})$ & $\hat{Q}_{yz} (\hat{J}_{y,s})$ &$\hat{Q}_{zx} (\hat{J}_{x,a})$ &$\hat{J}_y (\hat{J}_{y,a})$ & $\hat{J}_z$  &  $\hat{D}_{xy}$  & $\hat{Q}_{xy}$  & $\hat{Y}$ & $\hat{J}_{z,s}$ & $\hat{J}_{z,a}$ \\
 \hline\hline
$\hat{J}_x (\hat{J}_{x,s})$ & 0 &$  2i \hat{J}_{z,s}  $ & $-i \hat{Q}_{xy}$ & $i\hat{J}_z$ & $-i \hat{J}_y$ & $-i\hat{Q}_{yz}$ &$i\hat{Q}_{zx}$ & $-i\sqrt{3}\hat{Q}_{yz}$ & $-2i\hat{Q}_{yz}$ & $-i\hat{Q}_{yz}$\\ 
 \hline
 $\hat{Q}_{yz} (\hat{J}_{y,s})$ & $-2i \hat{J}_{z,s}$ & 0 & $-i\hat{J}_z$ & $-i\hat{Q}_{xy}$ & $i \hat{Q}_{zx} $ & $i \hat{J}_{x}$ & $i\hat{J}_y$ & $i\sqrt{3} \hat{J}_x$ & $i2\hat{J}_x$ & $i \hat{J}_{x}$  \\
 \hline
 $\hat{Q}_{zx} (\hat{J}_{x,a})$ & $i \hat{Q}_{xy}$ &  $i\hat{J}_z$ & 0 & $ 2i\hat{J}_{z,a}$ &  $-i\hat{Q}_{yz}$ &  $i \hat{J}_{y}$ &  $-i \hat{J}_{x}$ &  $-i\sqrt{3} \hat{J}_{y}$ &  $-i \hat{J}_{y}$ &  $-2i \hat{J}_{y}$\\
 \hline
 $\hat{J}_y (\hat{J}_{y,a})$ & $-i\hat{J}_z$ & $i\hat{Q}_{xy}$  & $ -2i\hat{J}_{z,a}$ & 0 & $i \hat{J}_x$ & $-i\hat{Q}_{zx}$ & $-i\hat{Q}_{yz}$ & $i\sqrt{3}\hat{Q}_{zx}$ & $i \hat{Q}_{zx}$ & $ 2i \hat{Q}_{zx}$ \\
 \hline
 $\hat{J}_z$ &  $i \hat{J}_y$ & $-i \hat{Q}_{zx} $ & $i\hat{Q}_{yz}$ & $-i \hat{J}_x$ & 0 & $ 2i \hat{Q}_{xy}$ & $- 2i \hat{D}_{xy}$ & 0 & $ i \hat{Q}_{xy}$& $-i \hat{Q}_{xy}$ \\
 \hline
  $\hat{D}_{xy}$& $i\hat{J}_{y,s}$ &  $-i \hat{J}_{x}$  &  $-i \hat{J}_{y}$ &  $i\hat{Q}_{zx}$ & $ -2i \hat{Q}_{xy}$ & 0 & $2i\hat{J}_z$ & 0 & 0 & 0 \\
 \hline
 $\hat{Q}_{xy}$ & $-i\hat{Q}_{zx}$ &  $-i\hat{J}_y$ &  $i \hat{J}_{x}$ & $i\hat{Q}_{yz}$ & $ 2i \hat{D}_{xy}$ & $-2i\hat{J}_z$ & 0 & 0 & $- i \hat{J}_z$& $ i \hat{J}_z$\\
 \hline
 $\hat{Y}$& $i\sqrt{3}\hat{J}_{y,s}$ & $-i\sqrt{3} \hat{J}_x$ & $i\sqrt{3} \hat{J}_{y}$ & $-i\sqrt{3}\hat{Q}_{zx}$ & 0 & 0 & 0 & 0 & 0 & 0\\
 \hline
 $\hat{J}_{z,s}$ &$2i\hat{J}_{y,s}$ & $-i2\hat{J}_x$ &  $i \hat{J}_{y}$& $-i \hat{Q}_{zx}$ & $-i \hat{Q}_{xy}$ & 0 & $ i \hat{J}_z$& 0 & 0 & 0\\
 \hline
 $\hat{J}_{z,a}$ & $i\hat{J}_{y,s}$ & $-i \hat{J}_{x}$  & $2i \hat{J}_{y}$ & $ -2i \hat{Q}_{zx}$ & $i \hat{Q}_{xy}$ & 0 & $ -i \hat{J}_z$& 0 & 0 & 0\\
 \hline
\end{tabular}
\end{table*}

In the case of spinor system we follow the same track of calculations as presented in the previous Appendix. First we rotate the spin-1 Hamiltonian (\ref{eq:spinorH}) around the $\hat{J}_{z,s}$ by $\pi/8$ angle. It is to locate the constant energy lines outgoing from a saddle fixed point along the $\hat{J}_{y,s}$ axis of the Bloch sphere in the symmetric subspace. However, this time the angle is two times smaller because commutation relations $[\hat{J}_{i,s}, \hat{J}_{j,s}]=i2 \hat{J}_{k,s} \epsilon_{ijk}$ contain the factor~$2$. After the rotation of Hamiltonian, one introduces the small parameter $\varepsilon=1/N$, transforming spin components into $\hat{h}_j=\sqrt{\varepsilon} \hat{J}_j$, $\hat{q}_j=\sqrt{\varepsilon} \hat{Q}_j$. The rotated and re-scaled Hamiltonian reads
\begin{align}
    \hat{H}&=-\frac{1}{\sqrt{\varepsilon}} \left[ 
    \frac{1}{2}\left(\hat{h}_{x,s} + \hat{h}_{y,s}\right)^2 
    + \left(\hat{h}_{y, a}{\rm cos}\frac{\pi}{8} + \hat{h}_{z, a}{\rm sin}\frac{\pi}{8}\right)^2 \right. \nonumber \\
    &+\left.
    \left(\hat{h}_{z}{\rm cos}\frac{\pi}{8} + \hat{q}_{xy}{\rm sin}\frac{\pi}{8}\right)^2
    +a \hat{n}_0 - a \hat{n}
    \right],
\end{align}
where $\hat{n}_0=\sqrt{\varepsilon}\hat{N}_0$, $ \hat{n} = \sqrt{\varepsilon}\hat{N}$, $a=2q/\varepsilon$ while the energy unit is $\sqrt{\varepsilon} |c'_2|/2$ and we introduced dimensionless time $\tau = \sqrt{\varepsilon} t |c'_2|/2\hbar$.

Equations of motion for expectation values $s_j=\langle \hat{h}_j\rangle$ and second order moments $\delta_{j,k}=\langle \hat{h}_j \hat{h}_k + \hat{h}_k \hat{h}_j\rangle  - 2 \langle \hat{h}_j \rangle \langle \hat{h}_k \rangle$ are much more complex as for bimodal condensates, but one can find the general structure quite similar.
The relevant for our purposes are
\begin{align}
    &\dot{s}_{zs}=-(\delta_{ys,ys} - \delta_{xs,xs}) 
    - \frac{\sqrt{2}}{4}(\delta_{ya,ya} - \delta_{xa,xa}) \label{eq:szs},\\
    &\dot{\delta}_{xs,xs}=-2 \delta_{xs,xs}s_{zs} - a \delta_{xs,ys},\label{eq:deltaxsxs} \\
    &\dot{\delta}_{ys,ys}=2 \delta_{ys,ys}s_{zs} + a \delta_{xs,ys},
    \label{eq:deltaysys}
\end{align}
for symmetric operators, and
\begin{align}
    &\dot{s}_{za}=-\frac{1}{2} (\delta_{ys,ys} - \delta_{xs,xs}) 
    - \frac{\sqrt{2}}{2}(\delta_{ya,ya} - \delta_{xa,xa}) \label{eq:sza},\\
    &\dot{\delta}_{xa,xa}=-\sqrt{2} \delta_{xa,xa}s_{za} - a \delta_{xa,ya},\label{eq:deltaxaxa} \\
    &\dot{\delta}_{ya,ya}=\sqrt{2} \delta_{ya,ya}s_{za} +a \delta_{xa,ya},
    \label{eq:deltayaya}
\end{align}
for anti-symmetric operators. 
In Table \ref{tab:tab1} we listed commutation relations useful to obtain (\ref{eq:szs}) - (\ref{eq:deltayaya}).

The initial spin coherent state $|0, \pi \rangle_{\rm S}$ gives the following non-zero initial values for $s_{z\sigma}(0)=-1/\sqrt{\varepsilon}$ and $\delta_{x\sigma,x\sigma}(0)=\delta_{y\sigma,y\sigma}(0)=1$ for $\sigma=s,a$. 
Equations for expectation values for first and second moments in the short-time expansion show that some terms appearing in the above equations are zero if their average values are initially zero, e.g. $\delta_{z,z}=0,\, \delta^q_{xy,xy}\simeq 0$. We did not put such terms in the final forms of Eqs. ~(\ref{eq:sx})~-~(\ref{eq:deltayy}).
The equations for symmetric and anti-symmetric operators are very similar to the one obtained for the bimodal system. There are two differences: (i) $s_{z\sigma}$ (with $\sigma=s,a$) in (\ref{eq:szs}) and (\ref{eq:sza}) play the role of $s_x$ in (\ref{eq:sx}) and (ii) symmetric and anti-symmetric subspaces are coupled to each other in (\ref{eq:szs}) and (\ref{eq:sza}). The coupling makes the scaling analysis a little more intricate. Taking both into account, one can use solutions from the previous Appendix and find 
\begin{align}
    s_{zs}(\tau)&=
    s_{zs}(0) - \nonumber \\
    &-\frac{{\rm cosh}(2s_{zs}(0)\tau)-1}{2s_{zs}(0)}
    -\frac{\sqrt{2}}{4}
    \frac{{\rm cosh}(\sqrt{2}s_{za}(0)\tau)-1}{\sqrt{2}s_{za}(0)},\\
    s_{za}(\tau)&=
    s_{za}(0) - \nonumber \\
    &-\frac{1}{2}\frac{{\rm cosh}(2s_{zs}(0)\tau)-1}{2s_{zs}(0)}
    -\frac{\sqrt{2}}{2}
    \frac{{\rm cosh}(\sqrt{2}s_{za}(0)\tau)-1}{\sqrt{2}s_{za}(0)}.
\end{align}
Note, the symmetric and anti-symmetric subspaces are coupled to each other and this has to be taken into account while explaining the scaling of $\delta_{x\sigma,x\sigma}$.

In order to explain the scaling of the first maximum, one needs to find a derivative of the variances in respect to time. Now, there are two equations for $\sigma = s$ and $\sigma=a$ that help to express relations among ${\rm cosh}$ having different arguments. The maximization of the QFI over the time provides the scaling of the maximum to be $|c'_2|t_{max}/\hbar={\rm ln}(8N/3)$ by keeping leading terms in $\varepsilon$. Finally, the value of the maximum of the QFI gives $F_{Q, {\rm S}}\simeq 4 \Delta^2 \hat{J}_{xs} \simeq \frac{16}{3}e^{-2/3} N^2\approx 2.8 N^2$ when considering the leading terms in $\varepsilon$.

\section{Explanation of the QFI stabilization after rotation in the long times limit for spinor system}\label{app:forspinors}

Here we use the same reasoning as presented in the main text concerning the bimodal system at the end of Section~\ref{sec:storage}.
We assume that the direction of interferometric rotation just before the rotation for spinor system is 
$\hat{\Lambda}_{{\rm S}, {\bf n}_{max}}(t_1^{-}) \approx \frac{\hat{J}_{x\sigma} \pm \hat{J}_{y \sigma}}{\sqrt{2}}$, with sign "+" for $\sigma=s$ and sign "-" for $\sigma=a$, and therefore
$F_{Q,{\rm S}}(t_1^{-})=4 \Delta^2 \left( \frac{\hat{J}_{x\sigma} \pm \hat{J}_{y \sigma}}{\sqrt{2}} \right)$, while after the rotation for $t\ge t_1^{+}$ one has
$\hat{\Lambda}_{{\rm S}, {\bf n}_{max}}(t) \approx \hat{J}_{ya} $ and  
$F_{Q,{\rm BI}}(t)=4 \Delta^2 \hat{J}_{ya} $.
It is a fairly good approximation, as demonstrated in Appendix~\ref{app:covariancematrixspinor} and in Figs.~\ref{fig:inter_norot-spinor} and \ref{fig:interfer_rot-spinor}.

The QFI after rotation for $t\leq t_1^{+}$ can be also written as
\begin{equation}
    F_{Q,{\rm S}}(t)=4 \left[-2N \frac{\langle\hat{H}_{\rm S}(t) \rangle}{c'_2} - 
    \langle \hat{J}^2_{xs}(t)  \rangle - \langle \hat{J}^2_{z}(t)  \rangle + q \langle \hat{N}_{0}(t)  \rangle \right],
\end{equation}
where we used (\ref{eq:spinorH}). Next, we note that the average energy is conserved after rotation, $\langle \hat{H}_{\rm S}(t) \rangle = \langle \hat{H}_{\rm S}(t_1^{+})\rangle$, while the average values of $\hat{J}^2_{xs}(t), \, \hat{J}^2_{z}(t), \, \hat{N}_{0}(t)$ are bounded from below by zero. These two properties lead to the inequality
\begin{equation}\label{eq:fq_explanation-spinor}
    F_{Q,{\rm S}}(t) \ge -8 N \langle \hat{H}_{\rm S}(t_1^{+}) \rangle.
\end{equation}
The energy of the spinor system after the rotation (\ref{eq:spinorrot}) with $\alpha_{\rm S}=\pi/4$ reads 
$\langle \hat{H}_{\rm S} (t_1^{+})\rangle = 
-\frac{1}{2N} \left[
\langle \hat{J}^2_{xs}(t_1^-) \rangle 
+ \frac{\langle \left( \hat{J}_{ya}(t_1^-)-\hat{J}_{xa}(t_1^-) \right)^2\rangle}{2}
+\frac{\langle Q_{xy}^2 \rangle}{2}  \right]$. 
Finally, one considers the latter in (\ref{eq:fq_explanation-spinor}) to show that
\begin{equation}
    F_{Q,{\rm S}}(t) \ge F_{Q,{\rm S}}(t^-_1),
\end{equation}
for $t\ge t_1^+$ as $\langle \hat{J}^2_{xs}(t_1^-) \rangle \ge 0$ and $\langle Q_{xy}^2 \rangle \ge 0$ as well.

\bibliography{bibliography}

\end{document}